\documentclass[12pt]{article}

\usepackage[english]{babel}
\usepackage{amsmath}
\usepackage{graphicx}
\usepackage{apacite}
\usepackage{url} 
\usepackage{caption,blkarray}
\usepackage{bm}
\usepackage[labelformat=simple]{subcaption}

\usepackage{algorithm}
\usepackage{algpseudocode}

\newtheorem{definition}{Definition}

\usepackage{tikz}
\usetikzlibrary{matrix,shapes,arrows,positioning,chains}

\usepackage{rotating}
\usepackage{tabularx}
\usepackage{makecell}
\usepackage{booktabs}
\usepackage{amsmath}
\usepackage{xcolor}
\usepackage{authblk}

\usepackage{lineno}

\usepackage{amsmath}
\usepackage{multirow}
\usepackage{epsfig, epstopdf}
\usepackage{colortbl}
\usepackage[backref=page]{hyperref}
\usepackage{lineno}
\usepackage{amssymb,amsmath,bm}
\usepackage{natbib}
\usepackage{setspace}
\usepackage{lscape}
\usepackage{longtable}
\usepackage{afterpage,latexsym,epsfig,float}
\bibpunct{(}{)}{;}{a}{}{,}

\RequirePackage{mathtools}
\allowdisplaybreaks
\usepackage[font=small,labelfont=bf]{caption}
\usepackage{hyperref}
\hypersetup{
     colorlinks   = true,
     citecolor    = blue
}

\newcommand{\mk}[1]{{\color{black}{#1}}}

\begin{document}

\def\spacingset#1{\renewcommand{\baselinestretch}%
{#1}\small\normalsize} \spacingset{1}




\title{\bf Efficient estimation for a smoothing thin plate spline in a two-dimensional space}
 \author[1]{Joaquin Cavieres}
\author[2]{Michael Karkulik}
\affil[1]{Georg-August-Universität Göttingen, Germany}
\affil[2]{Universidad Técnica Federico Santa María, Chile}
\maketitle

\bigskip
\begin{abstract}
Using a deterministic framework allows us to estimate a function with the purpose of interpolating data in spatial statistics. Radial basis functions 
are commonly used for scattered data interpolation in a d-dimensional space, however, interpolation problems have to deal with dense matrices. For the case of smoothing thin plate splines, we propose an efficient way to address this problem by compressing the dense matrix by an hierarchical matrix ($\mathcal{H}$-matrix) and using the conjugate gradient method to solve the linear system of equations. A simulation study was conducted to assess the effectiveness of the spatial interpolation method. The results indicated that employing an $\mathcal{H}$-matrix along with the conjugate gradient method allows for efficient computations while maintaining a minimal error. We also provide a sensitivity analysis that covers a range of smoothing and compression parameter values, along with a Monte Carlo simulation aimed at quantifying uncertainty in the approximated function. Lastly, we present a comparative study between the proposed approach and thin plate regression using the ``mgcv" package of the statistical software R. The comparison results demonstrate similar interpolation performance between the two methods. \end{abstract}

\noindent%
{\it Keywords:}  Smoothing thin plate spline, $\mathcal{H}$-matrix, spatial statistics, adaptive cross approximation.
\vfill

\newpage
\spacingset{1.5} 
\section{Introduction}
\label{section1}

In the last years, the increased availability of data from sensors, satellites, or monitoring stations sparked special interest in scattered data interpolation methods (\cite{xu2011bayesian}).
In this context, the main objective is to approximate a function $f:\mathbb{R}^d\rightarrow\mathbb{R}$ , given a set of sampled values
$f(\boldsymbol{x}_{1}),...., f(\boldsymbol{x}_{n})$ at a set of spatial sites $X = \{\boldsymbol{x}_{1},...., \boldsymbol{x}_{n}\} \subset \mathbb{R}^{d}$.

A commonly employed category of interpolants to approximate $f$ are radial basis functions (RBFs, \cite{scheuerer2013interpolation}). This technique was initially applied in the 1980s to problems in machine learning, and has since been used in different contexts such as partial differential equations, computer graphics, learning theory, medical imaging, environmental variables, cartography, and surface reconstruction, among others \citep{wendland2004scattered, chiles2009geostatistics}. RBFs are a powerful tool for problems of multivariate data interpolation
in arbitrary dimensions, such as required in spatial statistics \citep{wahba1990spline, lin2004spatial}. 

From a statistical perspective, several authors have introduced distinct methodologies involving RBFs. These encompass approaches like thin plate splines (TPS) and smoothing thin plate splines (STPS),  also used for spatial interpolation. Among these authors, \cite{wahba1990spline}, \cite{green1993nonparametric} and \cite{wang2011smoothing} offer an extensive read into smoothing splines. For example, \cite{wahba1975smoothing} shows how to choose the smoothing parameter for a periodic spline, \cite{hutchinson1985smoothing} presents a procedure to calculate the trace of the influence matrix for a polynomial smoothing spline of degree 2$m$-1, and \cite{craven1978smoothing} estimated the correct degree of smoothing using the Generalized Cross-Validation (GCV) method. An extension to the Bayesian context  can be found in \cite{wahba1983bayesian} and \cite{nychka1988bayesian}, who proposed an estimation of confident intervals for smoothing splines. Specifically, from the spatial interpolations applications, \cite{wahba1995smoothing} proposed an ANOVA analysis for an epidemiological study, \cite{trossman2011application} applied a TPS for comparative purposes in an oceanographic study and \cite{early2020smoothing} interpolated noisy GPS data using smoothing splines.  Another important contribution was made by \cite{wood2003thin}, wherein the author introduced an approximation of the splines matrix through truncated spectral decomposition. This allows us to get an efficient computation even when dealing with a large number of spatial observations (\cite{wood2017generalized}). On the other hand, the Kriging method (\cite{matheron1963principles}), is mostly used for statistical interpolation in dimension $d=2$. Both approximation methods, RBFs and Kriging, are similar and various authors have studied their connections \citep{scheuerer2013interpolation}. \mk{The computational cost for both methods
is cubic in the number of spatial observations, which is prohibitively high.}
To deal with this problem
different authors have proposed alternatives (\cite{lindgren2011explicit, bevilacqua2018geomodels, litvinenko2019likelihood}, and \cite{hristopulos2021stochastic}). 

From a deterministic perspective, the theory of \textit{Reproducing Kernel Hilbert Space} (RKHS) to interpolate spatial data (\cite{wahba1990spline}) is equivalent to the stochastic process assumed as random structure in the space \citep{berlinet2011reproducing}. Here, we consider the approximation of $f$ by a function $g: \mathbb{R}^{d} \rightarrow \mathbb{R}$
of the form
\begin{equation}\label{eq:1}
g(\boldsymbol{x}) =  \sum^{n}_{i = 1}c_{i}\Phi(\boldsymbol{x_i}, \boldsymbol{x}) +\sum^{q}_{\ell = 1}d_\ell p_\ell(\boldsymbol{x}),
\end{equation}
\mk{where $\Phi(\boldsymbol{x}, \boldsymbol{y}) = \phi(||\boldsymbol{x} - \boldsymbol{y}||_2)$ with $\phi:[0, \infty) \rightarrow \mathbb{R}$ radially symmetric and $||\cdot||_2$ the Euclidean norm in $\mathbb{R}^{d}$.}
Common choices of $\phi$ include conditionally semidefinite functions \citep[Chapter 8]{wendland2004scattered}, i.e., functions generating a definite interpolation problem
on the linear subspace given by $\sum_{i=1}^n c_i p(\boldsymbol{x}_i)=0$ for all polynomials
$p\in\mathcal{P}^d_m$ of degree less or equal to a certain number $m$ (depending on $\phi$) in dimension $d$.
Hence, the functions $\{p_{1},...., p_{q}\}$ in Equation \eqref{eq:1} are chosen to form a basis of $\mathcal{P}^d_m$. In other words, the interpolation problem
\begin{align}
\label{eq:2}
\begin{split}
\sum^{n}_{i = 1}c_{i}\Phi(\boldsymbol{x}_{i}, \boldsymbol{x}_{j}) \hspace{2mm} + \hspace{2mm} \sum^{q}_{\ell = 1}d_\ell p_\ell(\boldsymbol{x}_{j}) &= f(\boldsymbol{x}_j), \hspace{3mm} j = 1,...., n\\
\sum^{n}_{i = 1}c_{i}p_k(\boldsymbol{x}_{i}) &= 0,
\hspace{3mm} k = 1,...., q
\end{split}
\end{align}
has a unique solution $\boldsymbol{c} = (c_{1},...., c_{n})^{T} \in \mathbb{R}^{n}$ and $\boldsymbol{d} = (d_{1},...., d_{q})^{T} \in \mathbb{R}^{q}$.
Equations \eqref{eq:2} can be written as linear system
\begin{align}\label{eq:3}
  \begin{pmatrix}
    \boldsymbol{E} & \boldsymbol{P}\\
    \boldsymbol{P}^\top & \boldsymbol{0}
  \end{pmatrix}
  \begin{pmatrix}
    \boldsymbol{c} \\ \boldsymbol{d}
  \end{pmatrix}
  =
  \begin{pmatrix}
    \boldsymbol{f} \\ \boldsymbol{0}
  \end{pmatrix},
  \quad\text{ where }
  \boldsymbol{E}_{j,i} = \Phi(\boldsymbol{x}_{i}, \boldsymbol{x}_{j}) 
  \quad\text{ and \hspace{2mm}}
  \boldsymbol{P}_{j,i} = p_i(\boldsymbol{x}_j).
\end{align}
\mk{If $\Phi$ has local support, then the matrix $\boldsymbol{E}$ is sparsely populated
and system~\eqref{eq:3} can be solved in linear complexity \citep{wendland2006computational}.
Hence, there is no need for matrix compression techniques.
If $\Phi$ has non-local support then the matrix $\boldsymbol{E}$ is
fully populated, and this entails a high computational cost $\mathcal{O}(n^{3})$ for the solution of system~\eqref{eq:3}. As the rise of new technologies enables us to gather vast amounts of data, we need to address this high cost to make
computations viable \citep{iske2017hierarchical}.
Compression of densely populated matrices by certain data-sparse formats is a necessity for computations in various other areas, for example for the numerical solution of integral equations or, more recently, fractional partial differential equations. 
In this context, different
methods such as multipole expansions, panel clustering, wavelet compression, mosaic-skeleton
methods, adaptive cross approximation, and hybrid cross-approximation have been presented
(\cite{lyche2020numerical}, \cite{bini2019structured}).}
In particular, hierarchical matrices ($\mathcal{H}$-matrices) are a convenient and general algebraic format for
data-sparse compression, and in a nutshell they may be summarized as \textit{blockwise low-rank matrices} \citep{hackbusch2015,bebendorf2008hierarchical}. Although initially developed for application
to integral equations, an $\mathcal{H}$-matrix can be used in a much wider context as long
as the function generating the entries of the dense matrix ($\Phi$ in our case) is
\textit{asymptotically smooth} (\cite{iske2017hierarchical}). Hence, matrices in RBF interpolation and related problems
can be approximated well by an $\mathcal{H}$-matrix, allowing for a matrix-vector multiplication
of complexity $\mathcal{O}(n\log(n))$ and efficient iterative solution techniques, as witnessed
in \cite{iske2017hierarchical} and \cite{LoehndorfM_17}.

In this work, we consider the special case of smoothing thin plate spline (STPS) interpolation, including a regularization term measuring the smoothness of the fitted curve. We can rewrite this problem in
terms of a simple linear system of equations, where the dense splines matrix is compressed by an $\mathcal{H}$-matrix. To solve the linear system , we show how to rewrite it as a positive definite and symmetric problem, and present numerical experiments using the Conjugate Gradient method.

The paper is structured as follows: In Section \ref{section2}, we define the a thin plate spline (TPS) and a STPS. This section also describe general concepts related to the $\mathcal{H}$-matrix. In Section \ref{section3}, we present the results of a simulated study focuses on solving a linear equation system using three different methods: a direct solver, the Conjugate Gradient method, and, lastly, the Conjugate Gradient method combined with the $\mathcal{H}$-matrix. In addition, we assessed the quality of the approximated function, computed the estimation time, conducted sensitivity analyses for the penalty parameter and the parameters to build the $\mathcal{H}$-matrix, performed Monte Carlo simulations, and compared our approach with the thin plate spline regression (TPSR) proposed by \cite{wood2003thin}, also considering the associated uncertainty. In Section \ref{section4}, we conclude by discussing the results and the potential incorporation of hierarchical matrices into the \texttt{R} software.

\section{Methods}
\label{section2}

\subsection{Thin plate spline}
\label{subsec1}


A thin plate spline (TPS) is defined in terms of a conditionally semidefinite RBF (\cite{wendland2004scattered}) given by:
\vspace{-0.4cm}

\begin{equation}\label{eq:4}
\boldsymbol{\Phi}(\boldsymbol{x}) = \|\boldsymbol{x}\|^{2} \log \|\boldsymbol{x}\|, \hspace{2mm} \boldsymbol{x} \in \mathbb{R}^{d}
\end{equation}

where $\|\cdot\|$ is the Euclidean distance and the basis $\{p_1,p_2,p_3\}$ of
\vspace{-0.4cm}

\begin{equation*}
    \mathcal{P}^2_1 = \{ p:\mathbb{R}^2 \rightarrow \mathbb{R} \mid \exists a,b,c\in\mathbb{R}:\;
    p(\boldsymbol{x}) = a \boldsymbol{x}_1 + b\boldsymbol{x}_2 + c \}.
\end{equation*}

A function $g$ is called a TPS on a set
$\{ \boldsymbol{x}_1,\dots,\boldsymbol{x}_n\}\subset\mathbb{R}^2$ if it has the form

\begin{equation}\label{eq:5}
g(\boldsymbol{x}) = \sum^{n}_{i = 1}c_i\phi(||\boldsymbol{x} - \boldsymbol{x}_j||) +
\sum^3_{\ell = 1}d_{\ell}p_{\ell}(\boldsymbol{x}),
\end{equation}

with coefficients $c_1,\dots,c_n,d_1,d_2,d_3$ \citep{green1993nonparametric, cavieres2022thin}. If the vector of coefficients $\boldsymbol{c} = (c_1,\dots, c_n)^\top$ satisfies $\sum_{i=1}^n c_i p (\boldsymbol{x}_i) = 0$ for all $p\in\mathcal{P}^2_1$, then $g$ is called a natural TPS. Clearly, it is sufficient to impose the last identity only on a basis of $\mathcal{P}^2_1$, which leads to $\boldsymbol{P}^\top\boldsymbol{c} = \boldsymbol{0}$ for the matrix $\boldsymbol{P}\in\mathbb{R}^{n\times 3}$ given by $\boldsymbol{P}_{i,j} = p_j(\boldsymbol{x}_i)$.
\mk{While we stick to the case $d=2$ in order to simplify presentation and emphasize
applications to spatial statistics, we stress that TPS and the corresponding
compression by hierarchical matrices are not restricted to this case.}

\subsection{Smoothing thin plate spline (STPS)}
\label{subsec2}

A TPS provides an alternative option for scattered data approximation in the spatial domain, which does not consider a covariance function associated with spatial observations. In statistical modeling, considering the presence of inherent measurement uncertainty or ``noise" in collected data, we choose to approximate the function using a smoothing thin plate spline (STPS) \citep{wahba1990spline}. 

Let $g$ be an approximating function for $n$ observations $y_{i}$ at sites $\boldsymbol{x}_{i}$.
We can propose the following statistical model
\begin{equation}\label{eq:7}
y_{i} = g(\boldsymbol{x}_{i}) + \epsilon_{i},\quad i=1,\dots, n,
\end{equation}
where $\epsilon_{i}$ is a random error assumed as $\mathcal{N}(0,1)$. We can define the residual sum of penalized least squares terms by
\begin{equation}\label{eq:8}
S(g) = \sum^{n}_{i = 1} |y_{i} - g(\boldsymbol{x}_{i})|^{2} + \lambda J(g).
\end{equation}

The component $J(g)$ is a penalty function that measures the smoothness of $g$, and $\lambda$ is a penalty parameter that controls the balance between the data fit and smoothness. For a TPS in $\mathbb{R}^2$, the penalty function has the form

\begin{equation}\label{eq:9}
J(g) = \int_{\mathbb{R}^{2}} \bigg (\frac{\partial^2g}{\partial \boldsymbol{x}_1^2}\bigg )^{2}(\boldsymbol{x})
+2\bigg (\frac{\partial^2g}{\partial \boldsymbol{x}_1\partial\boldsymbol{x}_2}\bigg )^{2}(\boldsymbol{x})
+\bigg (\frac{\partial^2g}{\partial \boldsymbol{x}_2^2}\bigg )^{2}(\boldsymbol{x})\,d\boldsymbol{x}
\end{equation}

We aim to minimize $S(g)$ over the set of natural TPS, and according to~\citep[Thm.~7.1]{green1993nonparametric}, the analytical solution of Equation (\ref{eq:9}) is


\begin{equation*}
    J(g) = \boldsymbol{c}^\top \boldsymbol{E}\boldsymbol{c},
\end{equation*}

with $\boldsymbol{E}$ from Equation (\ref{eq:3}). Hence, the minimization of $S(g)$ over all natural TPS can be stated as

\begin{equation}\label{eq:10}
    \min_{\substack{\boldsymbol{c}\in\mathbb{R}^n,\boldsymbol{d}\in\mathbb{R}^3\\ \boldsymbol{P}^\top \boldsymbol{c}=\boldsymbol{0}}}
    ||\boldsymbol{y} - \boldsymbol{E}\boldsymbol{c} - \boldsymbol{P}\boldsymbol{d}||_2^{2} + \lambda\boldsymbol{c}^{T}\boldsymbol{E}\boldsymbol{c},
\end{equation}

It can be shown that if the sites $\{ \boldsymbol{x}_1,\dots,\boldsymbol{x}_n \}$ are not collinear, then Equation \eqref{eq:10} has a unique solution given by the linear system

\begin{align}\label{eq:13}
  \begin{pmatrix}
    \boldsymbol{E} + \lambda\boldsymbol{I} & \hspace{2mm} \boldsymbol{P}\\
    \boldsymbol{P}^\top & \hspace{2mm}\boldsymbol{0}
  \end{pmatrix}
  \begin{pmatrix}
    \boldsymbol{c} \\ \boldsymbol{d}
  \end{pmatrix}
  =
  \begin{pmatrix}
    \boldsymbol{y} \\ \boldsymbol{0}
  \end{pmatrix},
\end{align}

where $\boldsymbol{I}\in\mathbb{R}^{n\times n}$ is the identity matrix \citep[Sec.~7.1]{green1993nonparametric}.

\subsection{Hierarchical matrices ($\mathcal{H}$-matrix)}
\label{subsec4}
For a set of sites $\{\boldsymbol{x}_1,\dots, \boldsymbol{x}_n\}\subset\mathbb{R}^d$ we denote by $I=\{1,\dots,n\}$ the index set. A subset $\tau\subset I$ of indices is called a \textit{cluster}. For the above we can define the \textit{diameter} of a cluster $\tau$ as

\begin{equation*}
  \operatorname{diam}(\tau) = \max_{j,i\in\tau}
  \|\boldsymbol{x}_j-\boldsymbol{x}_i\|_2,
\end{equation*} 

and the distance of two clusters $\tau,\sigma$ as

\begin{equation*}
  \operatorname{dist}(\tau,\sigma) = 
  \min_{j\in\tau,i\in\sigma}\| \boldsymbol{x}_j - \boldsymbol{x}_i\|_2.
\end{equation*}

The quadratic cost in the number of indices of calculating diameter and distance can be avoided, e.g., by modified definitions of diameter and distance \citep{rjasanowSteinbach_07}, or the use of so-called \textit{bounding boxes} \citep{hackbusch2015}. \mk{For a parameter $\eta>0$ }
we introduce a so-called admissibility condition

\begin{equation*}
  \operatorname{adm}:
  \begin{cases}
  2^I\times 2^I &\rightarrow \{\texttt{true},\texttt{false}\}\\
  (\tau,\sigma) &\mapsto \min (\operatorname{diam}(\tau),
  \operatorname{diam}(\sigma)) < \eta
  \operatorname{dist}(\tau,\sigma).
  \end{cases}
\end{equation*}

A partition $\mathcal{P}$ of $I\times I$ so that all elements $p\in\mathcal{P}$ are of the form $p=\tau\times \sigma$ with clusters $\tau,\sigma\subset I$ is called \textit{block partition} and a $p$ of the above form is called \textit{block cluster}. For a constant $\sigma_{\rm small}>0$, a block partition is called
$\sigma_{\rm small}-$\textit{admissible}, if it can be split into two parts
$\mathcal{P} = \mathcal{P}_{\rm near}\cup \mathcal{P}_{\rm far}$ with $\operatorname{adm}(\tau,\sigma)=\texttt{true}$
for all $\tau\times\sigma\in\mathcal{P}_{\rm far}$ (the \textit{far field}) and either
$|\tau|\leq\sigma_{\rm small}$ or $|\sigma|\leq\sigma_{\rm small}$ for $\tau\times\sigma\in\mathcal{P}_{\rm near}$ (the \textit{near field}).

\begin{definition}\label{def:hmat} 
Let $\mathcal{P}$ be an admissible block partition of $I\times I$. We define the space of $\mathcal{H}$-matrices 
of rank $r\in\mathbb{N}$ as
\begin{align*}
    \mathcal{H}(\mathcal{P},r) :=
    \{B\in \mathbb{R}^{n\times n} \mid \forall \tau\times\sigma\in\mathcal{P}_{\rm far}\;
    \exists X\in\mathbb{R}^{|\tau|\times r},Y\in\mathbb{R}^{|\sigma|\times r}
    \text{ so that } B|_{\tau\times\sigma} = XY^\top \}
  \end{align*}
 
\end{definition}

For uniformly distributed sites $\{\boldsymbol{x}_1,\dots, \boldsymbol{x}_n\}\subset\mathbb{R}^d$, $\sigma_{\rm small}$-admissible block partitions can be created in complexity $\mathcal{O}(n\log(n))$,
for example by geometry-based construction using hierarchical partitions of index set $I$, so-called \textit{cluster trees} \citep{hackbusch2015}. Far-field blocks are admissible and big, while near-field blocks are small.
It can be shown that in this case, the storage requirement of a matrix in $\mathbf{A}_{\mathcal{H}}\in\mathcal{H}(\mathcal{P},r)$ is $\mathcal{O}(r n \log(n))$, while matrix-vector operations $\mathbf{x}\mapsto \mathbf{A}_{\mathcal{H}}\mathbf{x}$ can also be carried out in complexity $\mathcal{O}(rn\log(n))$.

Matrices $X,Y$ associated to far-field blocks in $\mathcal{P}_{\rm far}$ can be calculated by interpolation of the kernel function, which needs to be known explicitly \citep{iske2017hierarchical}. This approach can be shown to approximate a given matrix $\mathbf{A}_{j,i} = \Phi(\boldsymbol{s}_j,\boldsymbol{s}_i)$ via a hierarchical matrix $\mathbf{A}_{\mathcal{H}}$ exponentially in the rank $r$, if the kernel function $\Phi$ is \textit{asymptotically smooth} \citep{hackbusch2015},

\begin{equation*}
|\partial_{\boldsymbol{x}}^{\boldsymbol{\alpha}}\partial_{\boldsymbol{y}}^{\boldsymbol{\beta}}\Phi(\boldsymbol{x},\boldsymbol{y})|
  \leq C (\boldsymbol{\alpha}+\boldsymbol{\beta})! |\boldsymbol{\alpha}+\boldsymbol{\beta}| \gamma^{|\boldsymbol{\alpha}+\boldsymbol{\beta}|}
  |\boldsymbol{x}-\boldsymbol{y}|^{-|\boldsymbol{\alpha}|-|\boldsymbol{\beta}|-s},
\end{equation*}

for some independent constant $C>0$, where $\boldsymbol{\alpha},\boldsymbol{\beta}\in\mathbb{N}^d_0$ are multi-indices and $|\boldsymbol{\alpha}| = \alpha_1+\alpha_2$, $\boldsymbol{\alpha}! = \alpha_1\alpha_2$ for $\boldsymbol{\alpha} = (\alpha_1,\alpha_2)$. In particular,
the TPS kernel $\Phi$
is asymptotically smooth \citep{iske2017hierarchical,LoehndorfM_17}, and hence the matrix $\boldsymbol{E}$ (or parts of it) in the system
\eqref{eq:13} can be approximated exponentially in the rank $r$ \mk{by an $\mathcal{H}$-matrix.}
Black-box methods which do not require the explicit knowledge of $\Phi$ also exist, the most obvious one being the Singular Value
Decomposition (SVD). However, the SVD is not apt for large-scale computations due to its \mk{cubic} complexity. We employed the computationally
more convenient \textit{Adaptive Cross Approximation} method (ACA, \cite{bebendorf2008hierarchical}). \mk{Instead of a prescribed rank $r$, the ACA algorithm employs a prescribed error tolerance $\varepsilon>0$ for the approximation of the far-field blocks, which are then approximated by different ranks. Definition~\ref{def:hmat} easily
extends to this case.}

\subsection{Implementation}

Hierarchical matrices are still not implemented in the statistical software \texttt{R}. As a remedy, we link the C\texttt{++} library \texttt{Htool} \citep{marchand2020schwarz} to \texttt{R} via \texttt{Rcpp} \citep{eddelbuettel2011rcpp} and \texttt{RcppArmadillo} \citep{eddelbuettel2014rcpparmadillo}. With these latter two libraries we conducted all of the numerical experiments
presented in Section (\ref{section3}) directly in \texttt{R}. To solve the linear system we use the Conjugate Gradient method \citep{hestenes1952methods} along with Schur complement \citep{zhang2006schur} (Appendix \ref{section5}).

\section{Results}
\label{section3}

\subsection{Matrix compression via an $\mathcal{H}$-matrix}

\begin{figure}[ht!]
  \begin{subfigure}[b]{0.5\linewidth}
    \centering
    \includegraphics[width=6.2cm, height=5.6cm]{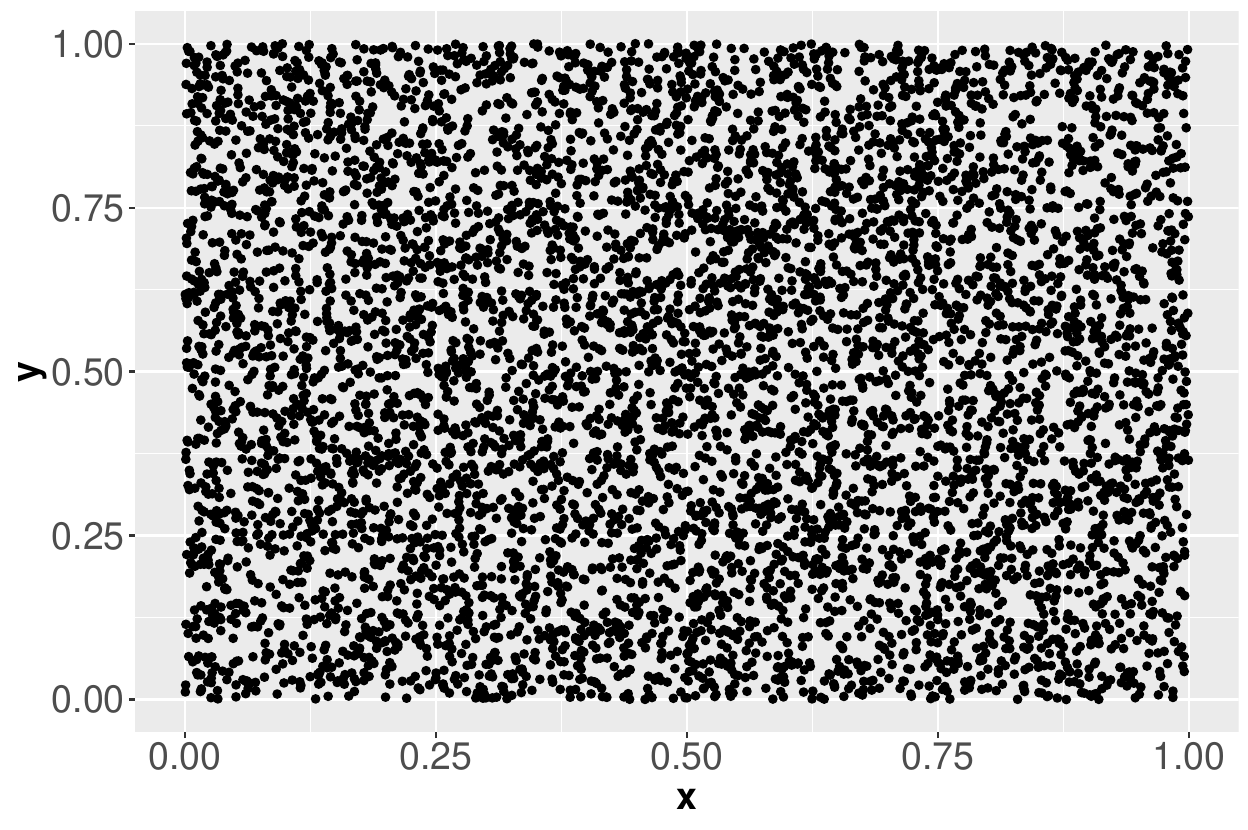} 
    \caption{\footnotesize{Unstructured points of size $80^{2}$}} 
    \end{subfigure}
     \hspace*{\fill}
  \begin{subfigure}[b]{0.5\linewidth}
    \centering
    \includegraphics[width=10cm, height=6cm]{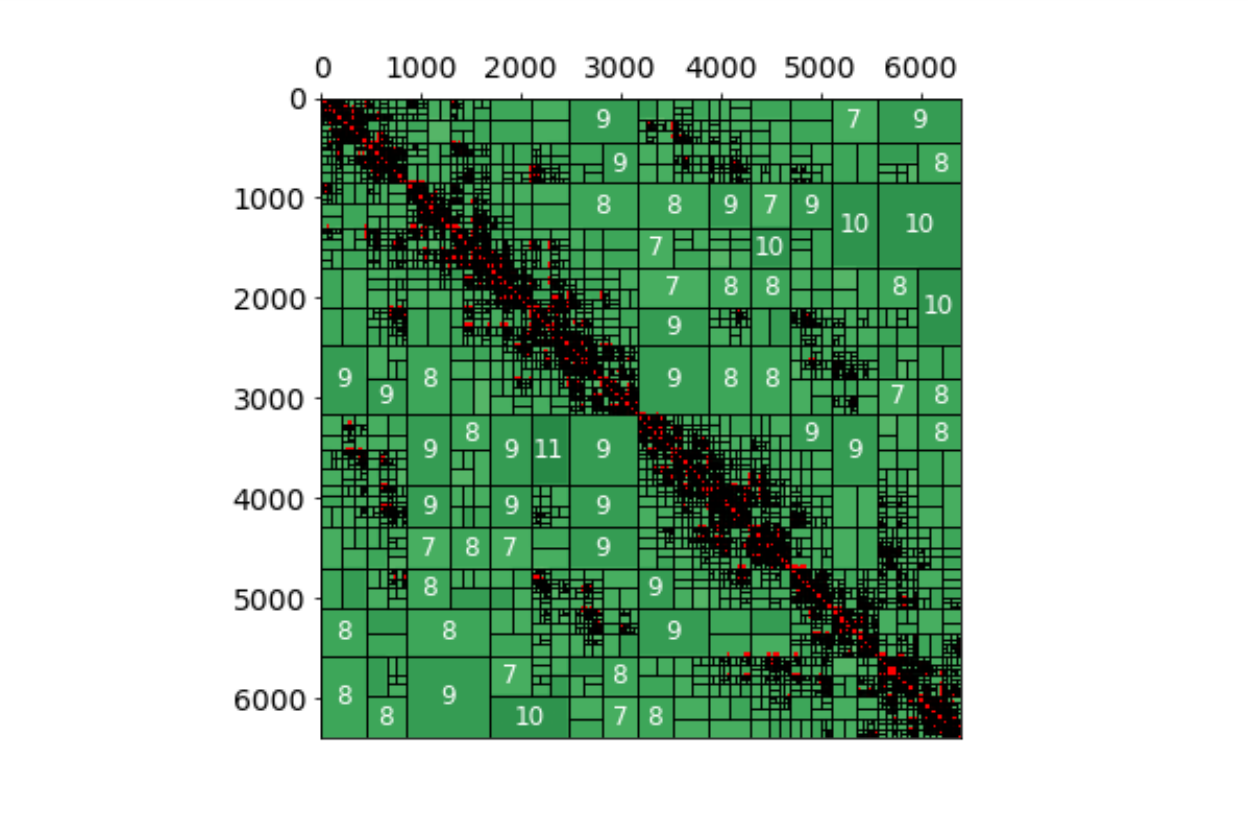} 
    \caption{\footnotesize{$\mathcal{H}$-matrix approximation of the dense matrix $\boldsymbol{E}$}}
 \end{subfigure}
  \vspace{4ex}
  \caption{\label{fig:plot1} Unstructured points of size $80^{2}$ (left side) and its respective $\mathcal{H}$-matrix (right side).}
 \end{figure}

First, we consider a grid of size $80^2$, $\lambda=1$, $\varepsilon = 0.0001$, $\eta = 2$ as parameters for the $\mathcal{H}-$matrix compression. In Figure (\ref{fig:plot1}), we present the grid and the corresponding compressed $\mathcal{H}$-matrix indicating the local rank on admissible blocks as determined by the ACA algorithm according to the given tolerance $\varepsilon$. Here the number of points in the grid were chosen to have a better visualization of the $\mathcal{H}$-matrix and how the block matrices are distributed on it. Besides, this number of points will allow us to determine if, while the number of observations increases, then the computational cost has a linear growth for M3. 


We studied the quality of the approximation (Subsection \ref{sub1}), the computational cost for three different approaches (Subsection \ref{sub2}), and the performance of the method under varying parameter settings (Subsection \ref{sub3}) considering the following:

\begin{enumerate}
    \item[\textbf{M1}]: solving \eqref{eq:13} with a direct solver provided by the \texttt{solver()} function in \texttt{R},
    \item[\textbf{M2}]: writing \eqref{eq:13} equivalently as symmetric positive system
    \eqref{eq:16} and using the Conjugate Gradient method,
    \item[\textbf{M3}]: writing \eqref{eq:13} equivalently as symmetric positive system
    \eqref{eq:16} and using the Conjugate Gradient method,
    approximating the dense matrix $\boldsymbol{E}_{11}$ from~\eqref{eq:16} by an $\mathcal{H}$-matrix, given the tolerance $\varepsilon$ for ACA and the admissibility coefficient $\eta$ as compression parameters.
\end{enumerate}

Experiments were done for a sequence of random grids generated by a two-dimensional uniform distribution on the square $(0,1)\times (0,1)$.
The data $\boldsymbol{y}_i = f(\boldsymbol{x}_i)$, were generated by the Franke's function;
\newpage

\begin{align*}
    f(x,y) = \frac{3}{4} e^{(-\frac{1}{4}(9x - 2)^{2} + (9y - 2)^{2})} + \frac{3}{4} e^{(-\frac{1}{49}(9x + 1)^{2} - (\frac{1}{10})(9y + 1)^{2})} + &\\
    \frac{1}{2} e^{(-\frac{1}{4}(9x - 7)^{2} + (9y - 3)^{2})} - & \frac{1}{2} e^{(-\frac{1}{5}(9x - 4)^{2} + (9y - 7)^{2})},
\end{align*}

which is a standard test function for two-dimensional spatial data fitting \citep{fasshauer2007meshfree}, and the spatial statistical model can be represented as

\begin{equation}\label{eq:smooth}
y_{i} = f(\boldsymbol{x}_{i}) + \varepsilon_{i}, \hspace{4mm} \text{for} \hspace{4mm} i = 1, \ldots, n, \hspace{4mm} \text{and} \hspace{4mm} \varepsilon_{i} \sim N(0, \sigma^{2}\boldsymbol{I})
\end{equation}

Naturally, we expect the computational time of the method M1 to scale cubically in the number of sites. For method M2 we expect $mn^2$, $m$ being the number of iterations of the conjugate gradient algorithm. However, the computational time of the method M3 is expected to scale like $mrn\log(n)$.

\subsection{Quality of the approximation}\label{sub1}

We assume that method M1 is \textit{computationally exact}, as it is carried out using a direct solver without matrix compression.
Let $f_j$, $j=1,2,3$ denote the solutions of the three different methods, and
$\boldsymbol{f}$, respectively $\boldsymbol{f}_j$, be the vector given by the evaluations of $f$, respectively $f_j$, on a grid of size $80^2$. In Table \ref{table:table1}, we compare the computational error of the three different methods by,

\begin{equation*}
    \texttt{comp-err}_j = \| \boldsymbol{f}_1 - \boldsymbol{f}_j \|_2,
\end{equation*}

and obtain the expected result that $\texttt{comp-err}_3$ ist slightly bigger due to compression.

\begin{table}[ht!]
\centering
\caption{Computational error of the methods M2 and M3 compared with M1 for a grid of size $80^2$.}
\label{table:table1}\textbf{}
\addtolength{\tabcolsep}{8pt} 
\begin{tabular}{ccc}
\hline
Error  & j=2 &  j=3\\
\hline
$\texttt{comp-err}_j$ & 2.6e-06 & 0.19\\
\hline
\end{tabular}
\end{table}

\begin{figure}[H]
    \centering
    \includegraphics[width=9cm, height=8.5cm]{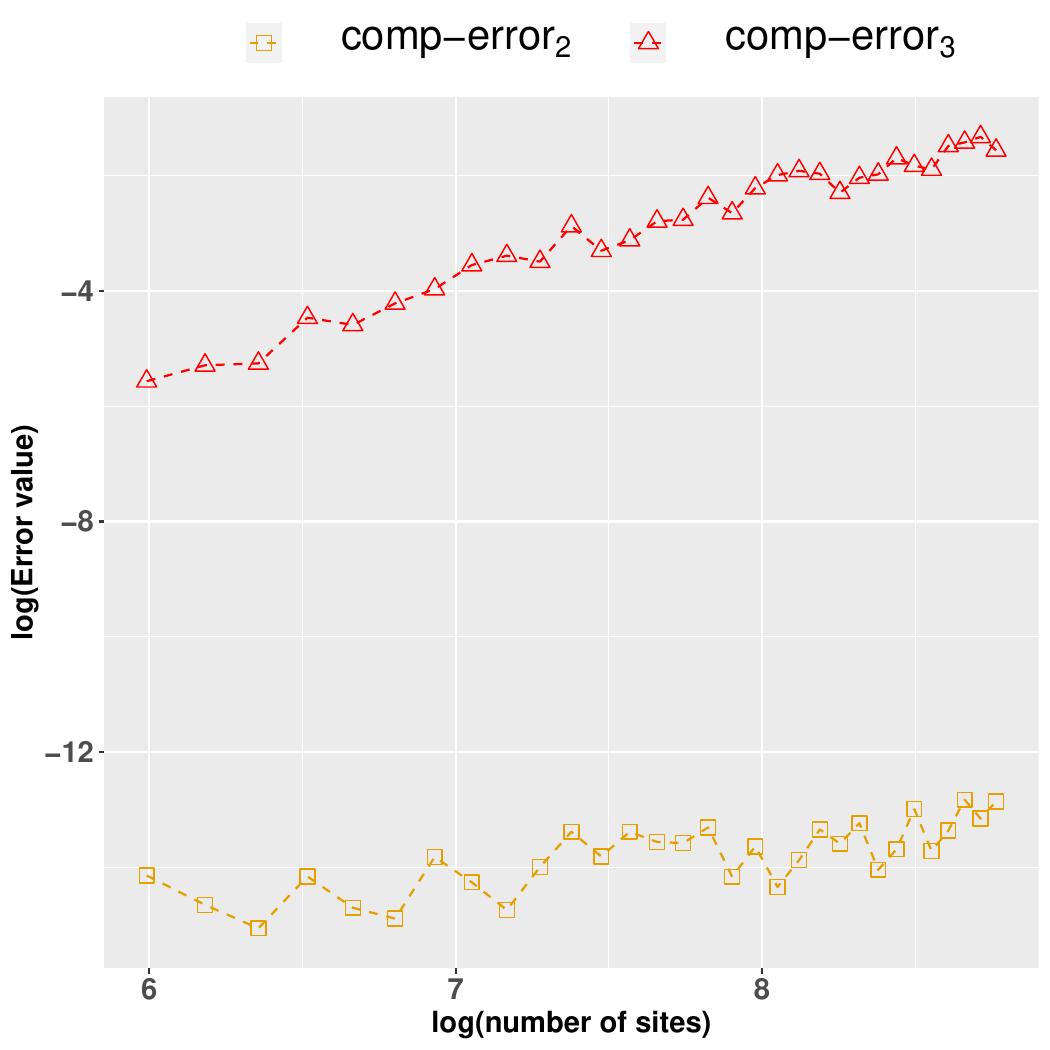}
\caption{\texttt{comp-error} of estimation of the coefficients by the M1, M2 and M3 method for different sizes of the grid in logarithmic scale}  
\label{fig:plot3}
\end{figure}

We also considered a sequence of grids with size ranging from $20^2$ to $80^2$, generated by a uniform distribution on the square $(0,1)\times (0,1)$ to compare the $\texttt{comp-err}_j$, on logarithmic scales (Figure \ref{fig:plot3}). 

Although the $\texttt{comp-err}_3$ increases as the number of sites is increasing as well, the scale of the error remains small.

To assess the interpolation performance of the three methods, we estimated the functions using M1, M2, and M3 on a grid of size $80^2$. Subsequently, we interpolated these functions onto a new grid of size $40^2$. For comparison purposes, we also applied the Franke's function to the spatial coordinates to obtain the ``real" values ($\boldsymbol{f}_{\text{real}}$), and then compare the interpolated function with those values using the Root Mean Square Error (\texttt{RMSE}) criterion,

\begin{equation*}
    \texttt{rmse}_j = \sqrt{\frac{1}{n} \sum_{i=1}^{n} (\boldsymbol{f}_{\text{real}} - \boldsymbol{f}_j)^2}
\end{equation*}

to measure the deviation of the STPS.


\begin{table}[ht!]
\centering
\caption{\texttt{rmse} computed for M1, M2, and M3. The ``real" values were obtained by applying the Franke's function to a grid of size $40^2$.}
\label{table:table2}\textbf{}
\addtolength{\tabcolsep}{8pt} 
\begin{tabular}{cccc}
\hline
  & M1 &  M2 & M3\\
\hline
$\texttt{rmse}_j$ & 0.01  & 0.01  & 0.01 \\
\hline
\end{tabular}
\end{table}

\begin{figure}[H]
\begin{minipage}{.47\linewidth}
\centering
\subfloat[]{\label{main:a}\includegraphics[scale=.30]{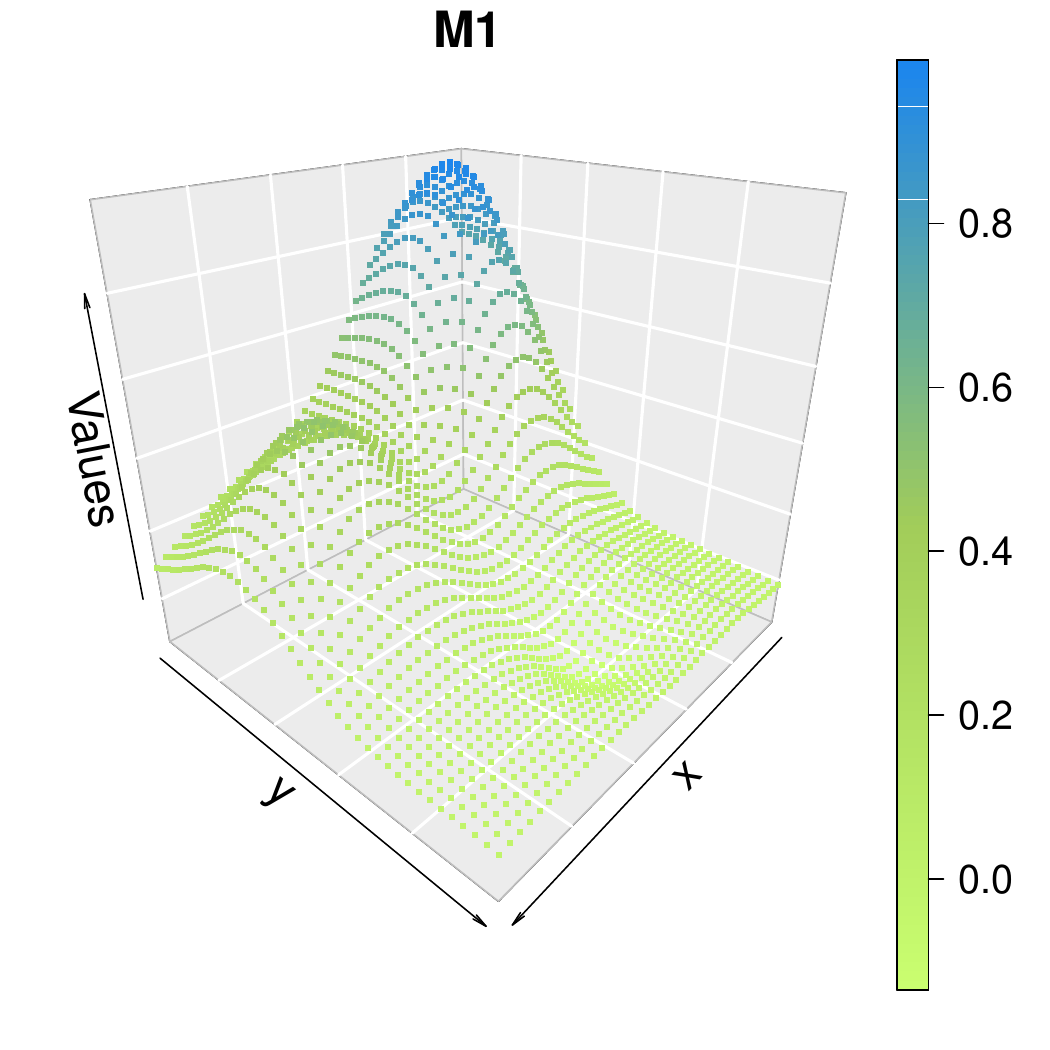}}
\end{minipage}%
\begin{minipage}{.47\linewidth}
\centering
\subfloat[]{\label{main:b}\includegraphics[scale=.30]{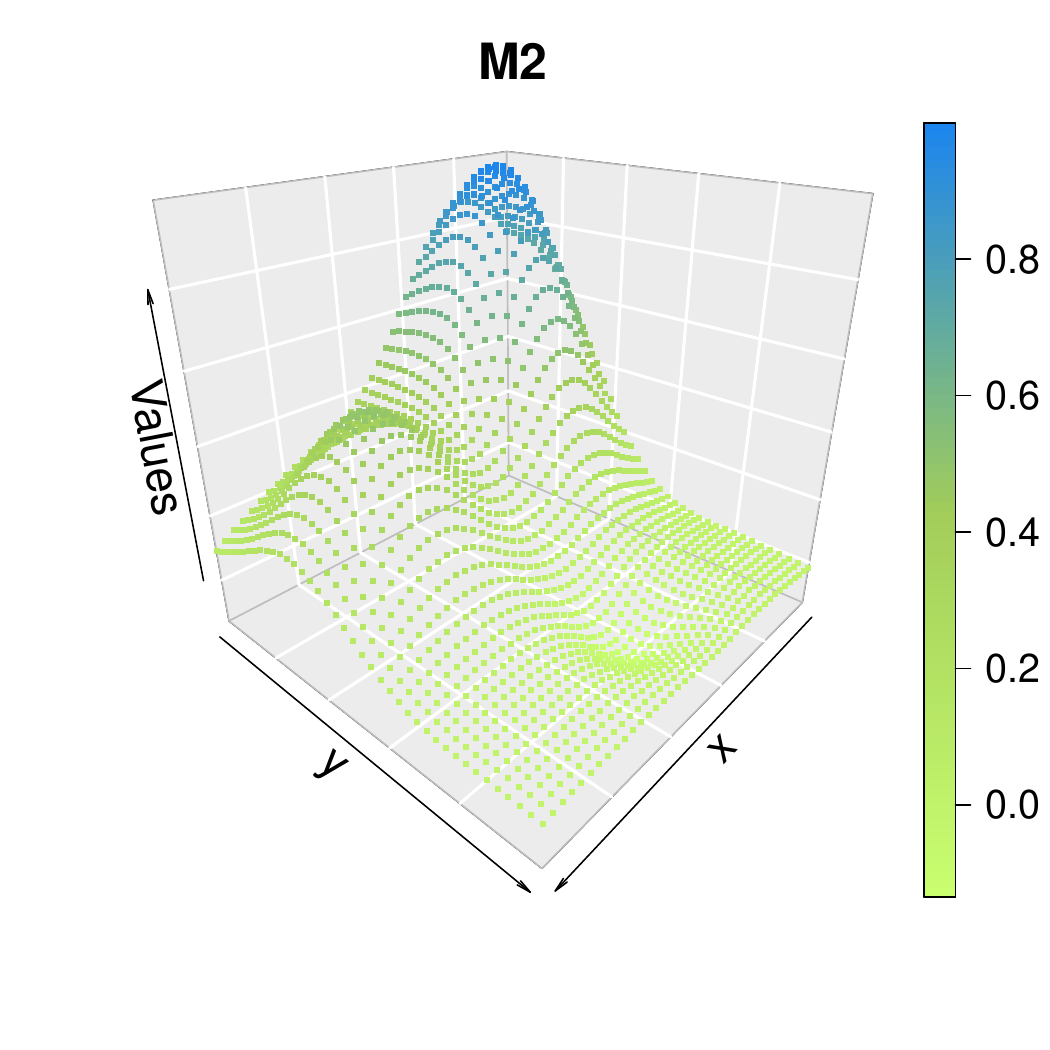}}
\end{minipage}\par\medskip
\centering
\subfloat[]{\label{main:c}\includegraphics[scale=.30]{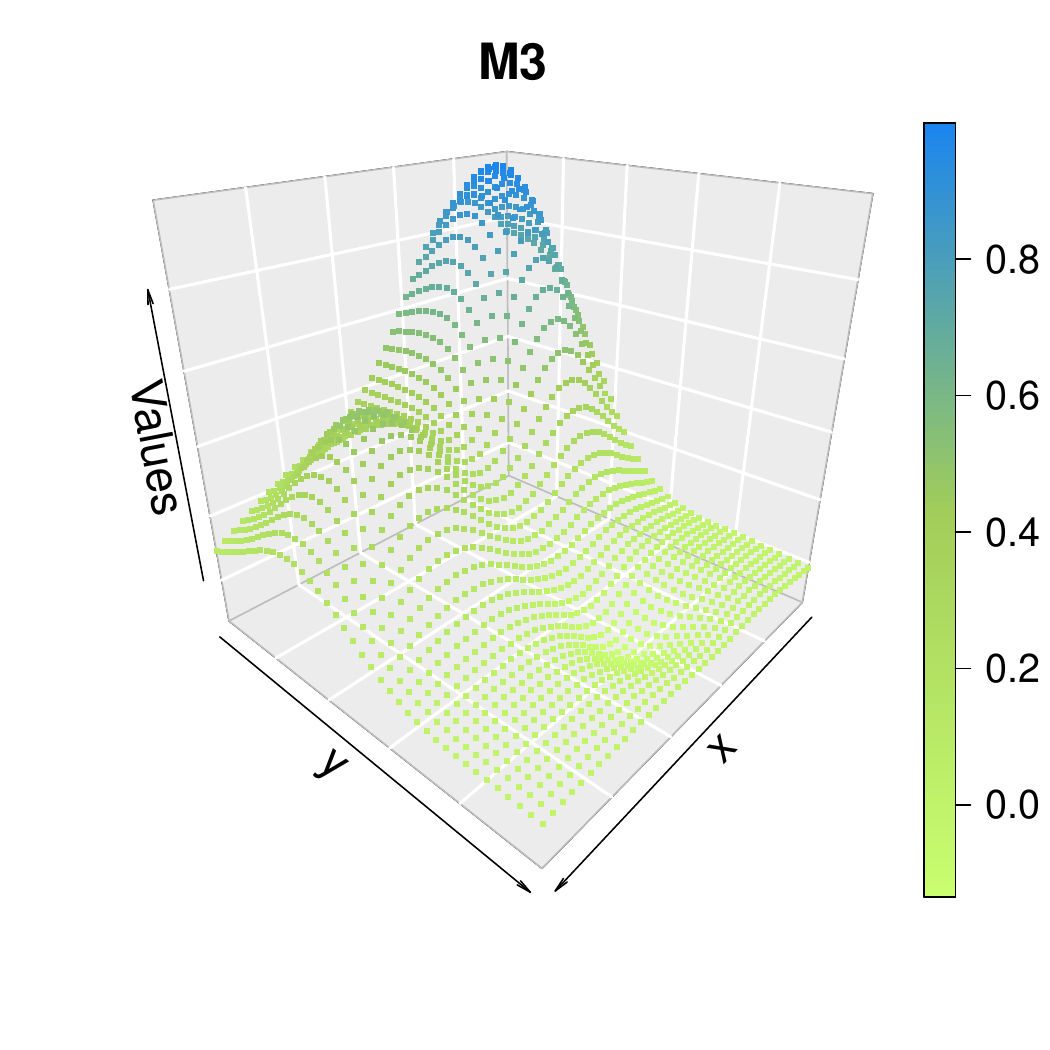}}
\caption{Spatial interpolation in a grid of $40^2$ based on the functions calculated from the grid of size $80^2$ by M1 (a), M2 (b) and M3 (c) respectively.}
\label{fig:plot2}
\end{figure}

The three methods have similar structures, indicating that though the $\mathcal{H}$-matrix does not have the complete data, it gives us a good approximation of the interpolated function (Figure \ref{fig:plot2}). Additionally, the \texttt{rmse} value is the same for all three methods (Table \ref{table:table2}). 

\subsection{Computational efficiency}\label{sub2}

\begin{figure}[H]
  \centering
    \includegraphics[width=9cm, height=8.5cm]{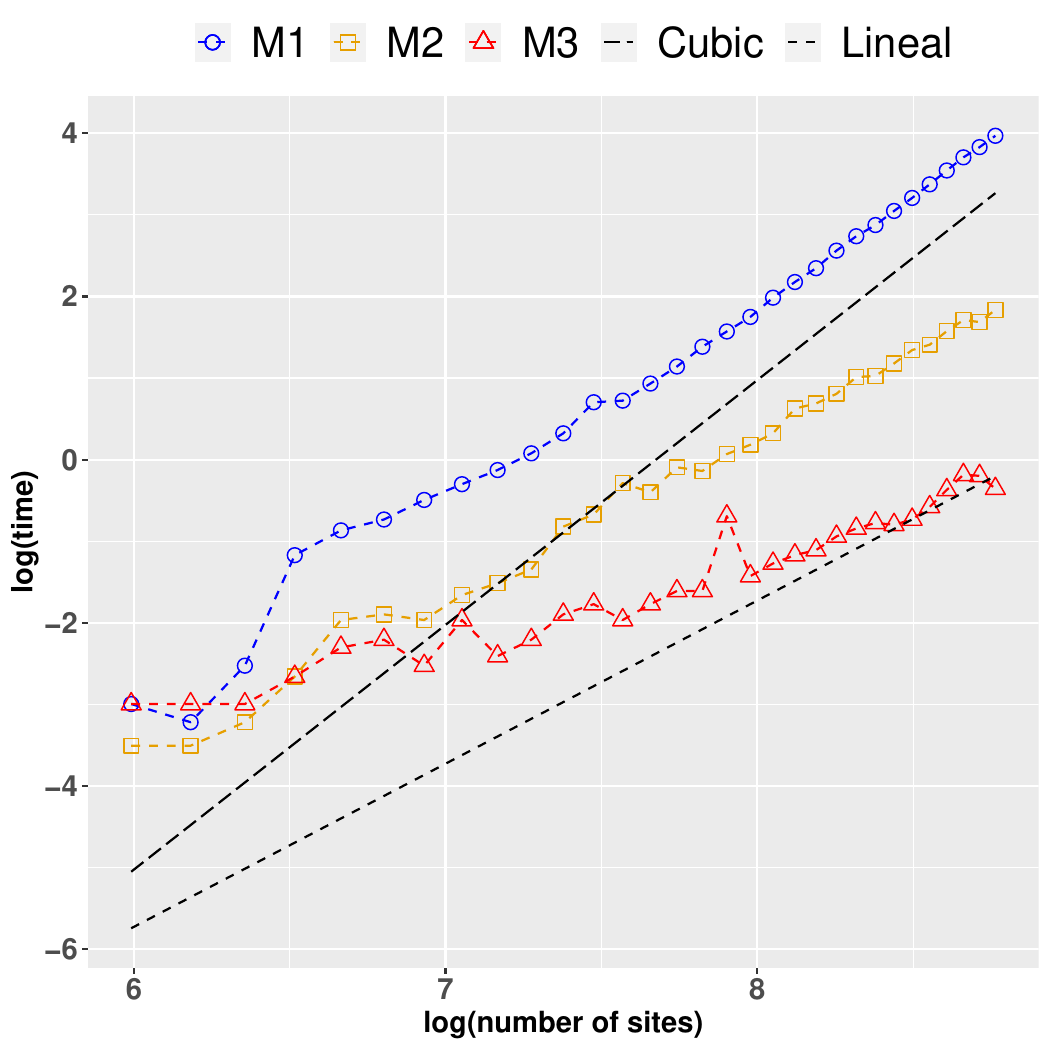}
\caption{Time of estimation of the coefficients by the M1, M2 and M3 method for different sizes of the grid in logarithmic scale.}  
\label{fig:plot4}
\end{figure}

We plot the computational time for solving the linear system using the three methods mentioned before (also in logarithmic scales) for a sequence of grids with size ranging from $20^2$ to $80^2$ as well. As expected, the computational time of the method M1 scales cubically in the number of sites. We use a fixed tolerance for the conjugate gradient algorithm, and hence, as expected, the method M2 scales quadratically in the number of sites, while for method M3 we observe practically a linear growth (Figure \ref{fig:plot4}).

\begin{figure}[ht!]
  \begin{subfigure}{0.49\linewidth}
    \centering
    \includegraphics[width=7.5cm, height=7.5cm]{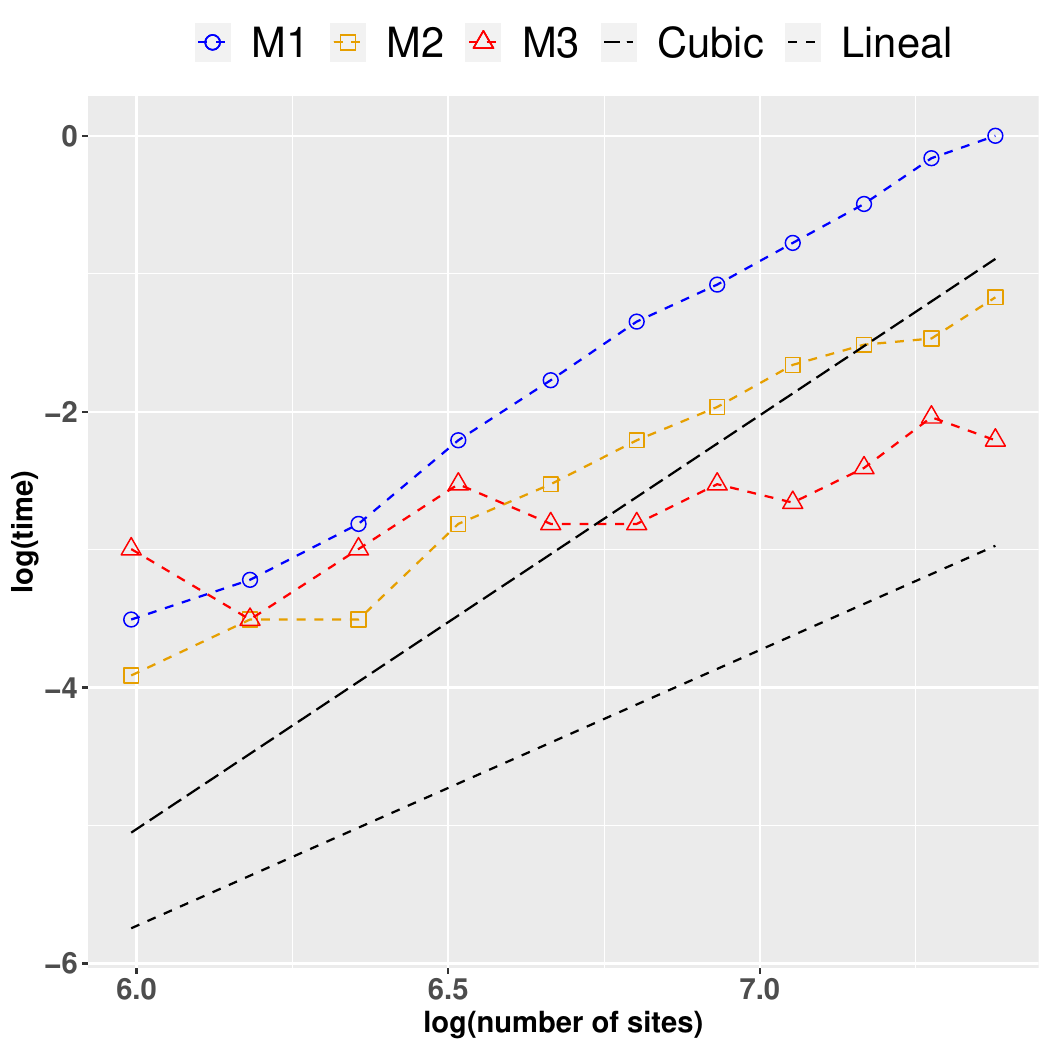} 
     \end{subfigure}
  \begin{subfigure}{0.49\linewidth}
    \centering
    \includegraphics[width=7.5cm, height=7.5cm]{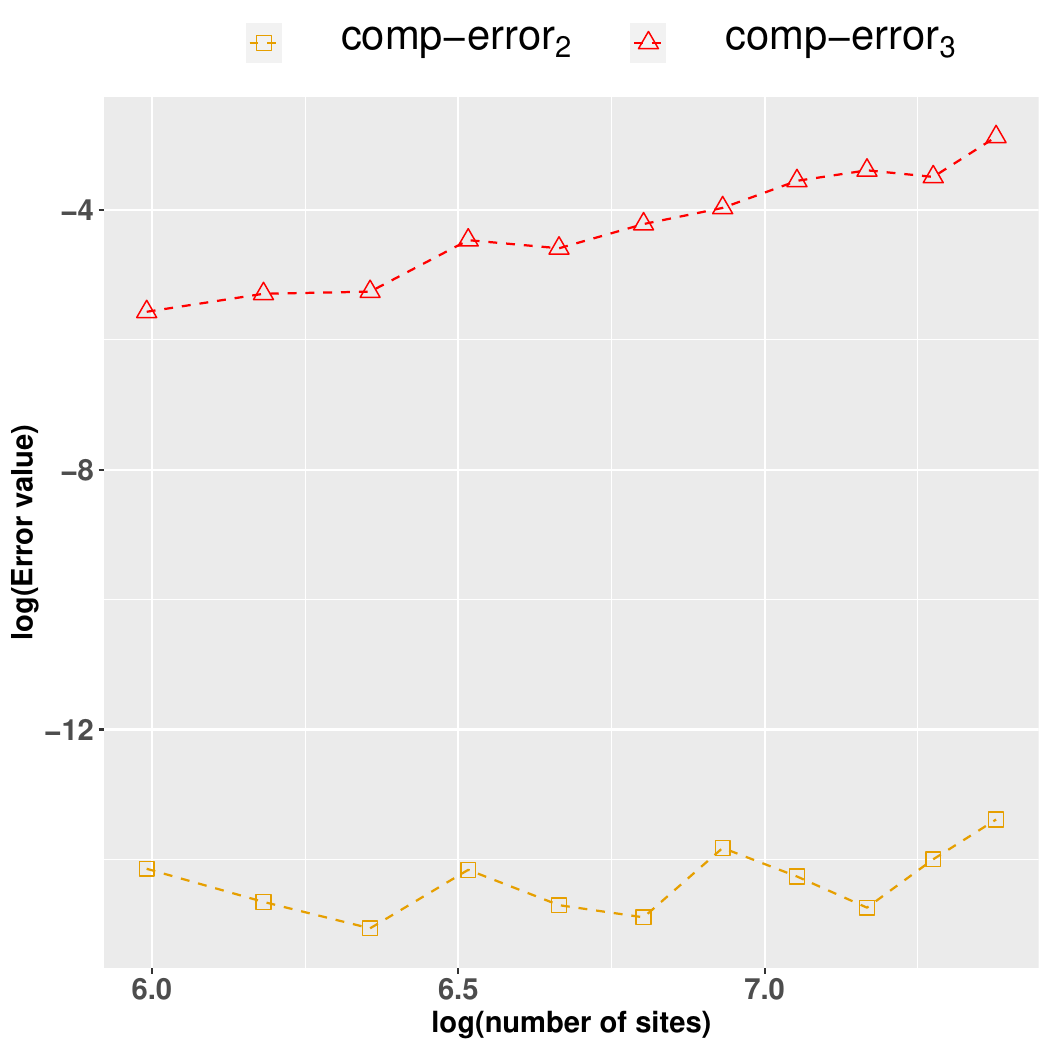} 
  \end{subfigure}
  \vspace{4ex}
  \caption{\label{fig:plot5} Time of estimation (left) and \texttt{comp-error} (right) for M1, M2 and M3 computed in a small sequences of grids, both are represented on logarithmic scales.}
 \end{figure}

We also show a comparison in computational time and the \texttt{comp-error} for a small sequence of grids from 20$^{2}$ to 40$^{2}$. This analysis considered the base case with $\varepsilon = 0.0001$ and $\eta = 2$. After increasing the number of observations, here also the $\mathcal{H}$-matrix (M3) allows us a fast computation compared with other methods (Figure \ref{fig:plot5}), and the error in the last grid (40 $\times$ 40) remains relatively minimal in contrast to M1 (Table \ref{table:table3}).

\vspace{0.5cm}

\begin{table}[ht!]
\centering
\caption{Computational error of the methods M2 and M3 compared with M1 for a grid of 40 $\times$ 40.}
\label{table:table3}\textbf{}
\addtolength{\tabcolsep}{8pt} 
\begin{tabular}{ccc}
\hline
Error  & j=2 &  j=3\\
\hline
$\texttt{comp-err}_j$ & 1.55e-06 & 0.05\\
\hline
\end{tabular}
\end{table}

\newpage

\subsection{Sensitivity analysis}\label{sub3}

Various scenarios were assessed to determine the effects of different compression parameters in method M3 for spatial interpolation. The procedure outlined in subsection \ref{sub1} was employed for these evaluations.

\begin{figure}[H]
  \centering
  \begin{minipage}{.48\linewidth}
    \centering
      {\includegraphics[width=7.5cm, height=6.8cm]{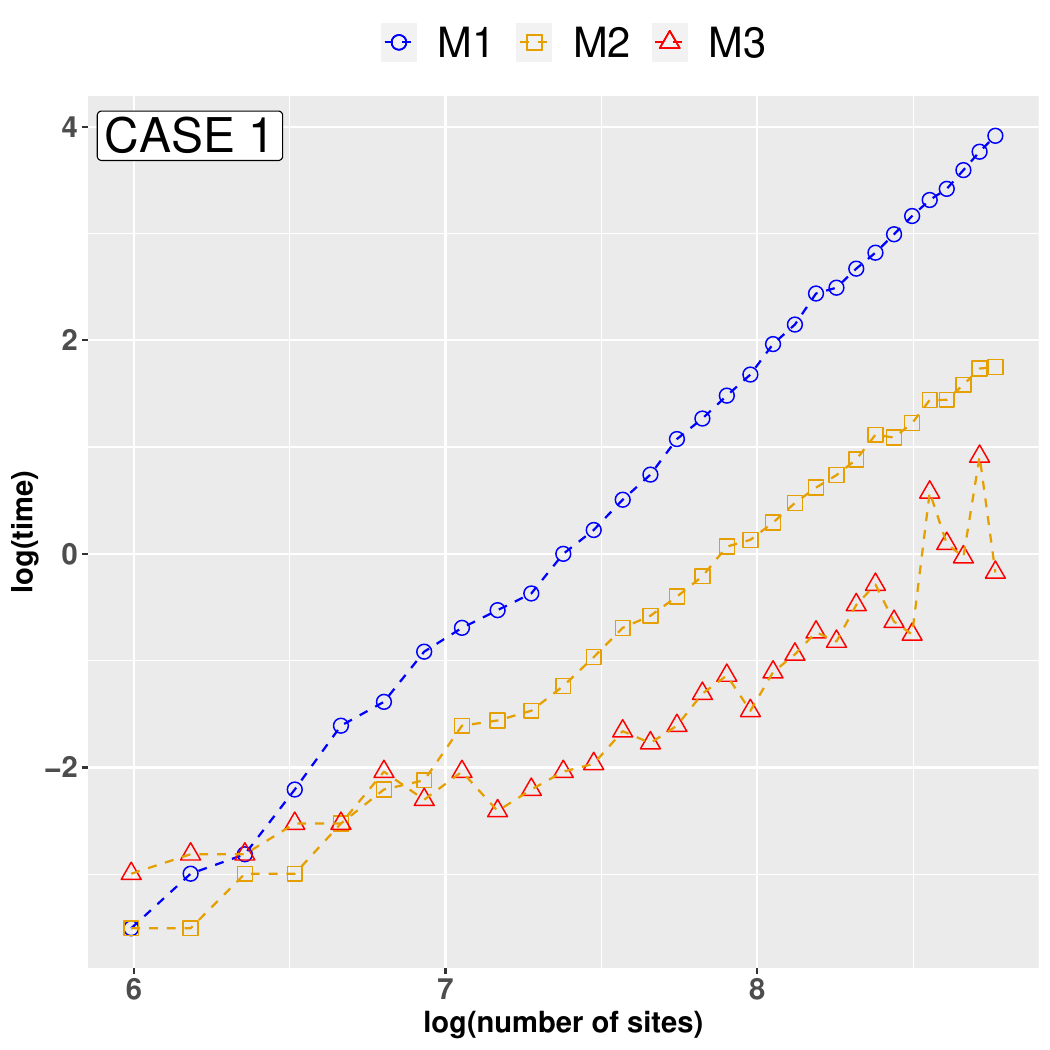}}
      {\includegraphics[width=7.5cm, height=6.8cm]{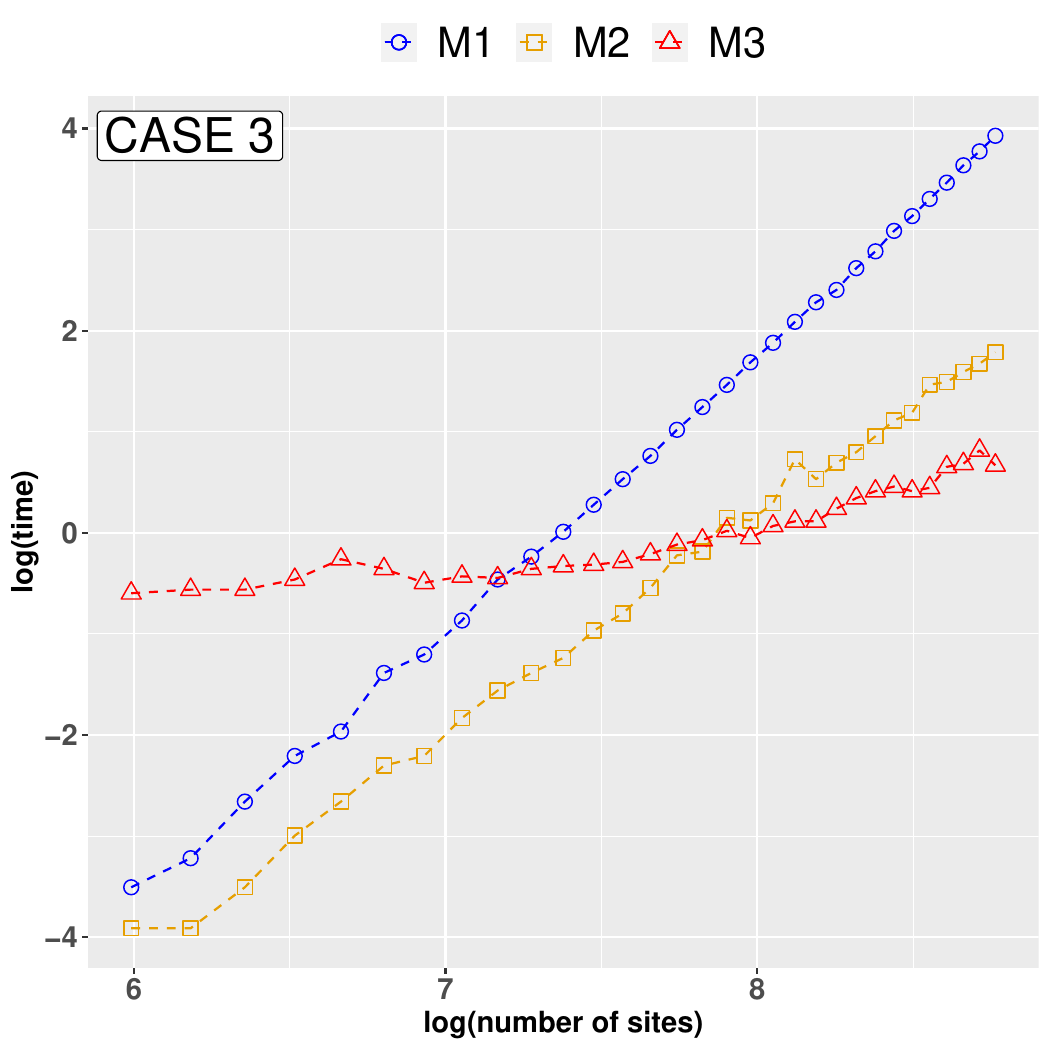}}
      \end{minipage}\quad
  \begin{minipage}{.48\linewidth}
    \centering
      {\includegraphics[width=7.5cm, height=6.8cm]{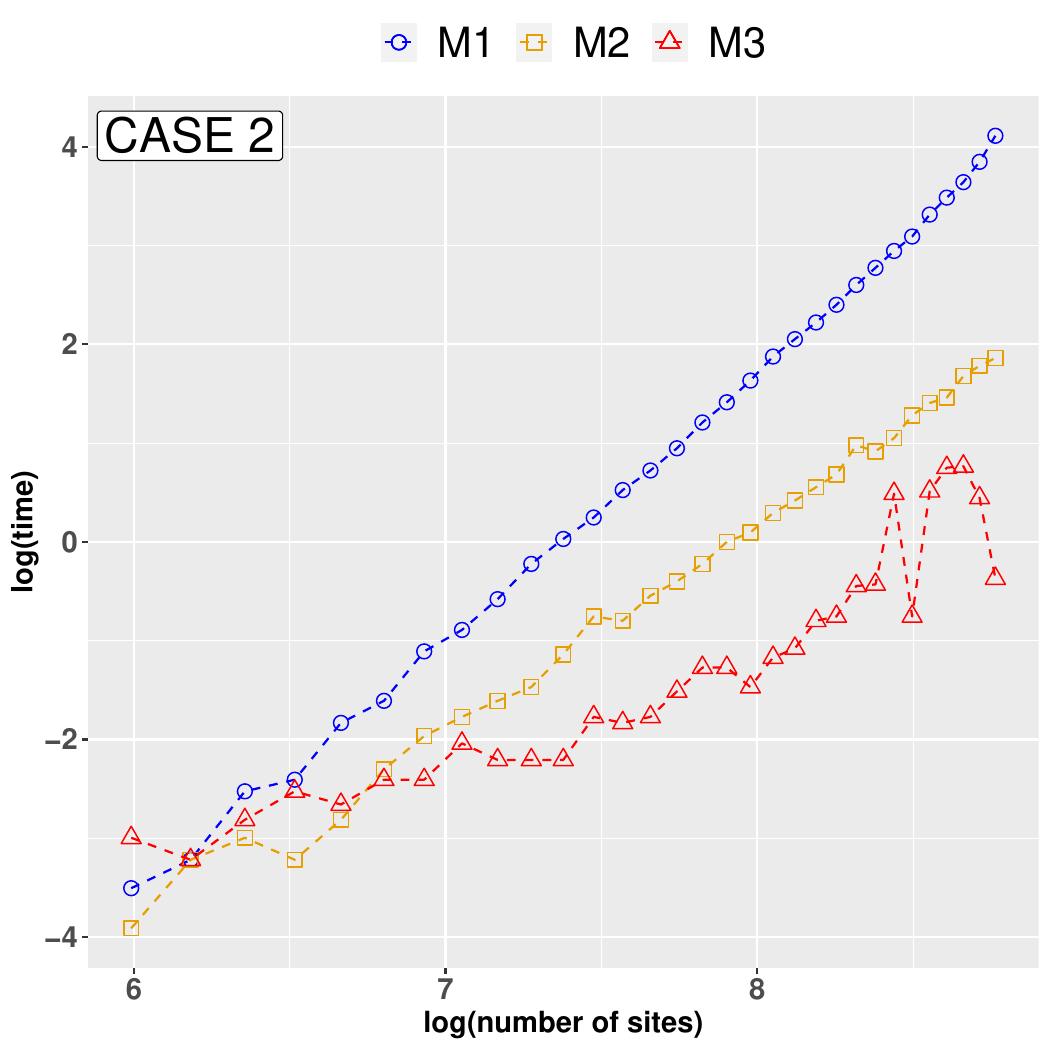}}
      {\includegraphics[width=7.5cm, height=6.8cm]{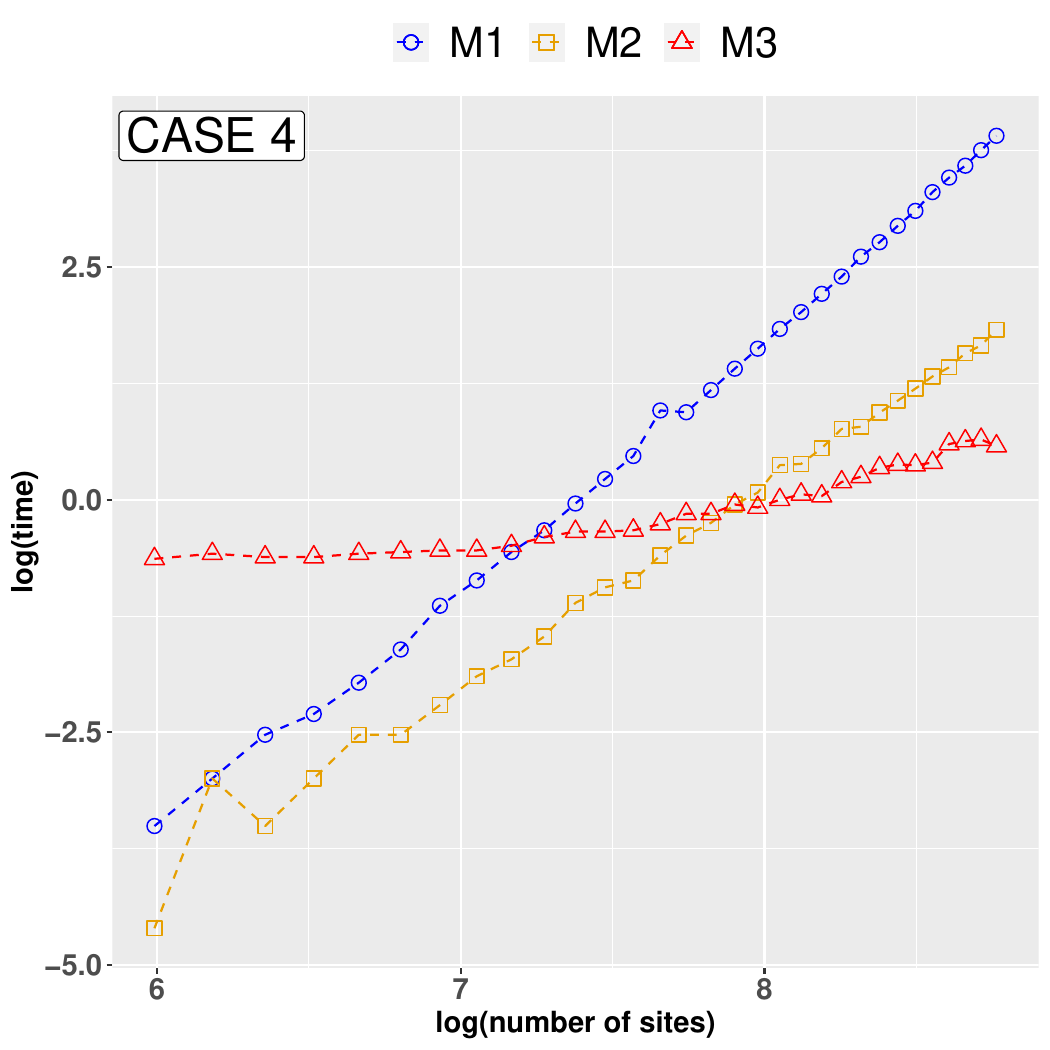}}
      \end{minipage}
\caption{Times of estimation for M1, M2 and M3 using different parametrizations of the $\mathcal{H}$-matrix}
\label{fig:plot6}
\end{figure}
\begin{figure}[H]
  \centering
  \begin{minipage}{.48\linewidth}
    \centering
      {\includegraphics[width=7.5cm, height=6.8cm]{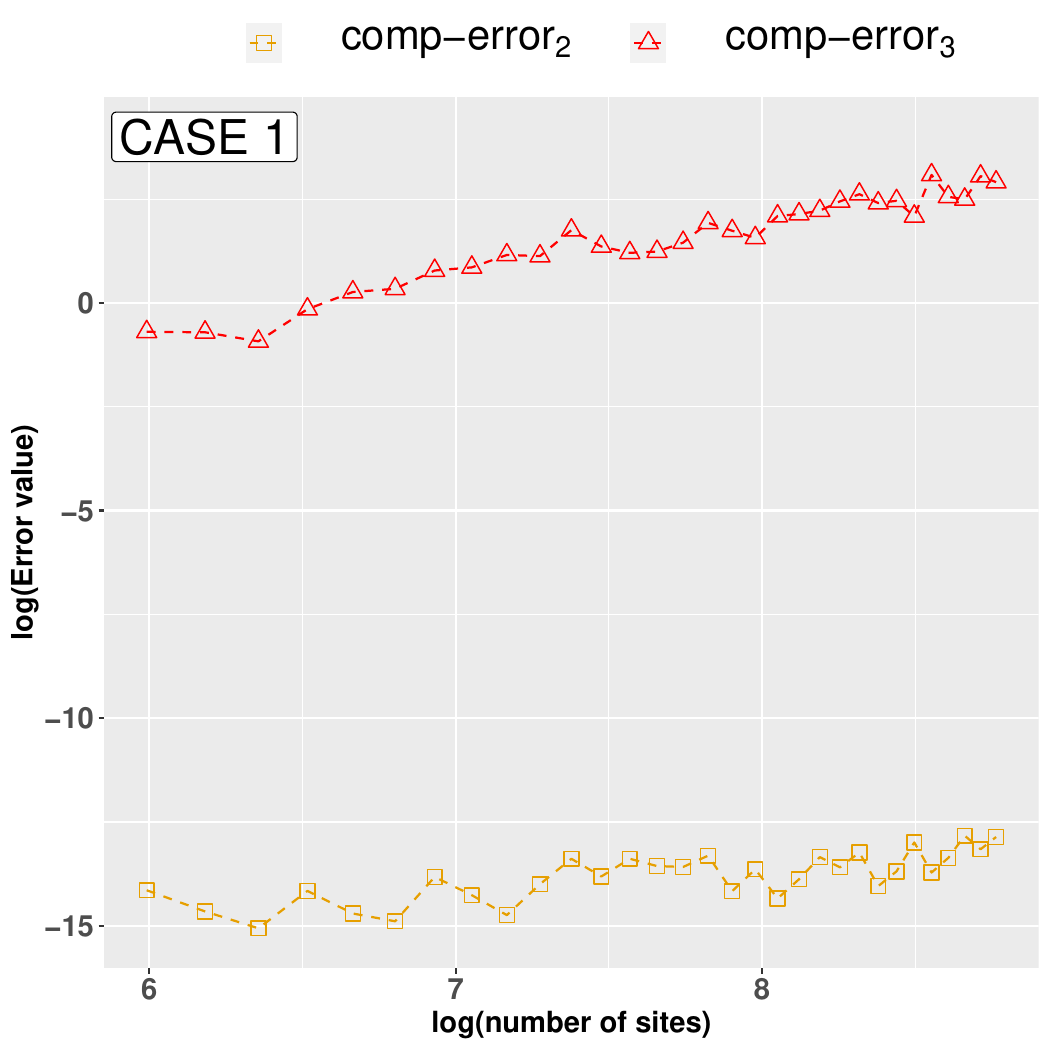}}
      {\includegraphics[width=7.5cm, height=6.8cm]{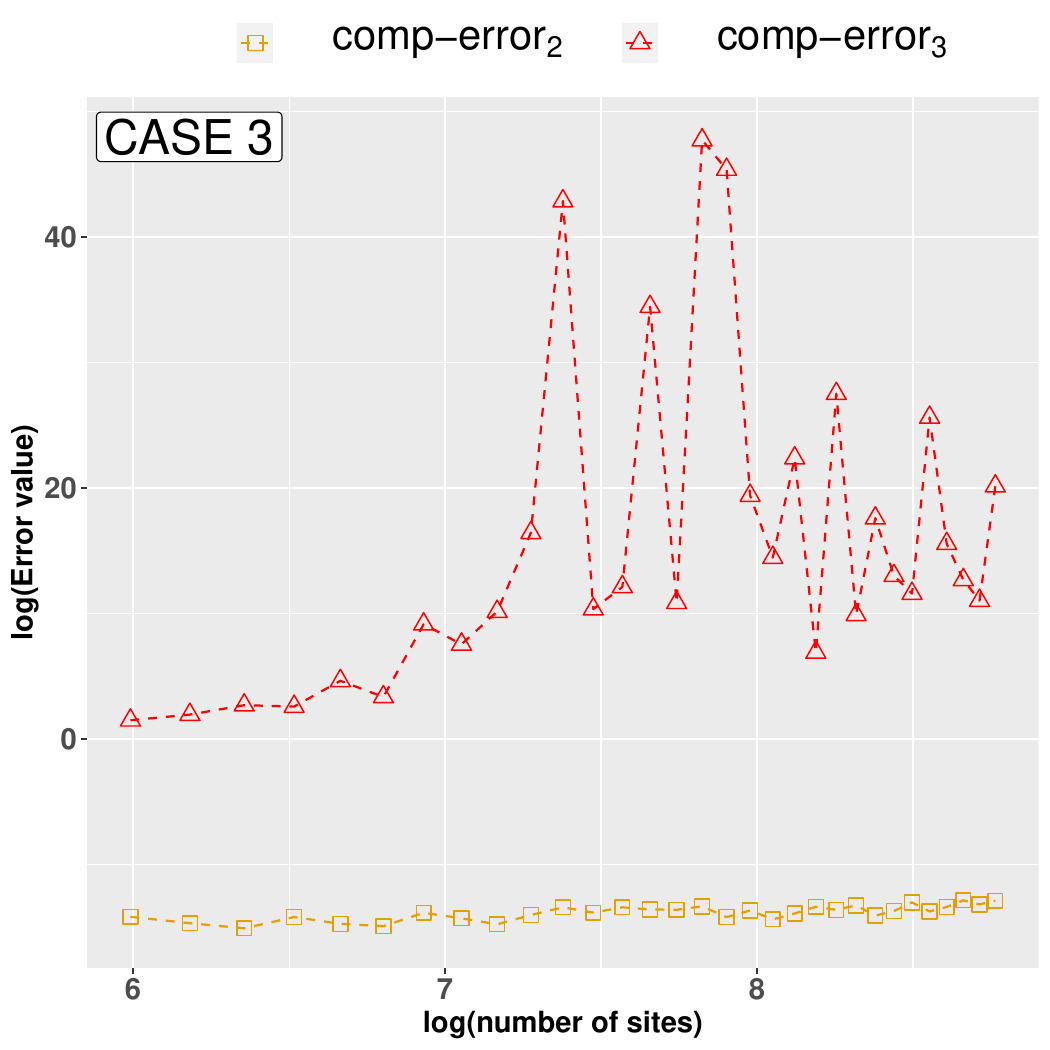}}
      \end{minipage}\quad
  \begin{minipage}{.48\linewidth}
    \centering
      {\includegraphics[width=7.5cm, height=6.8cm]{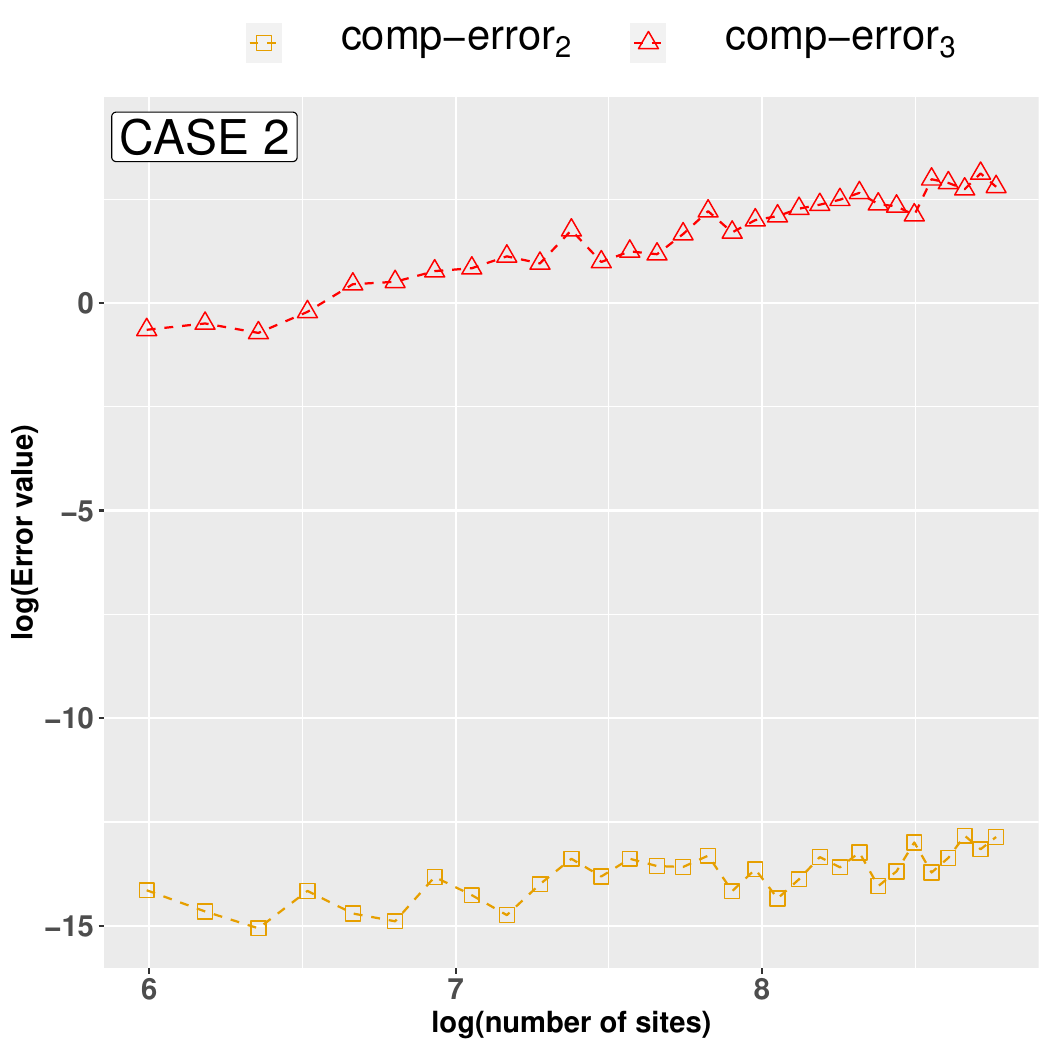}}
      {\includegraphics[width=7.5cm, height=6.8cm]{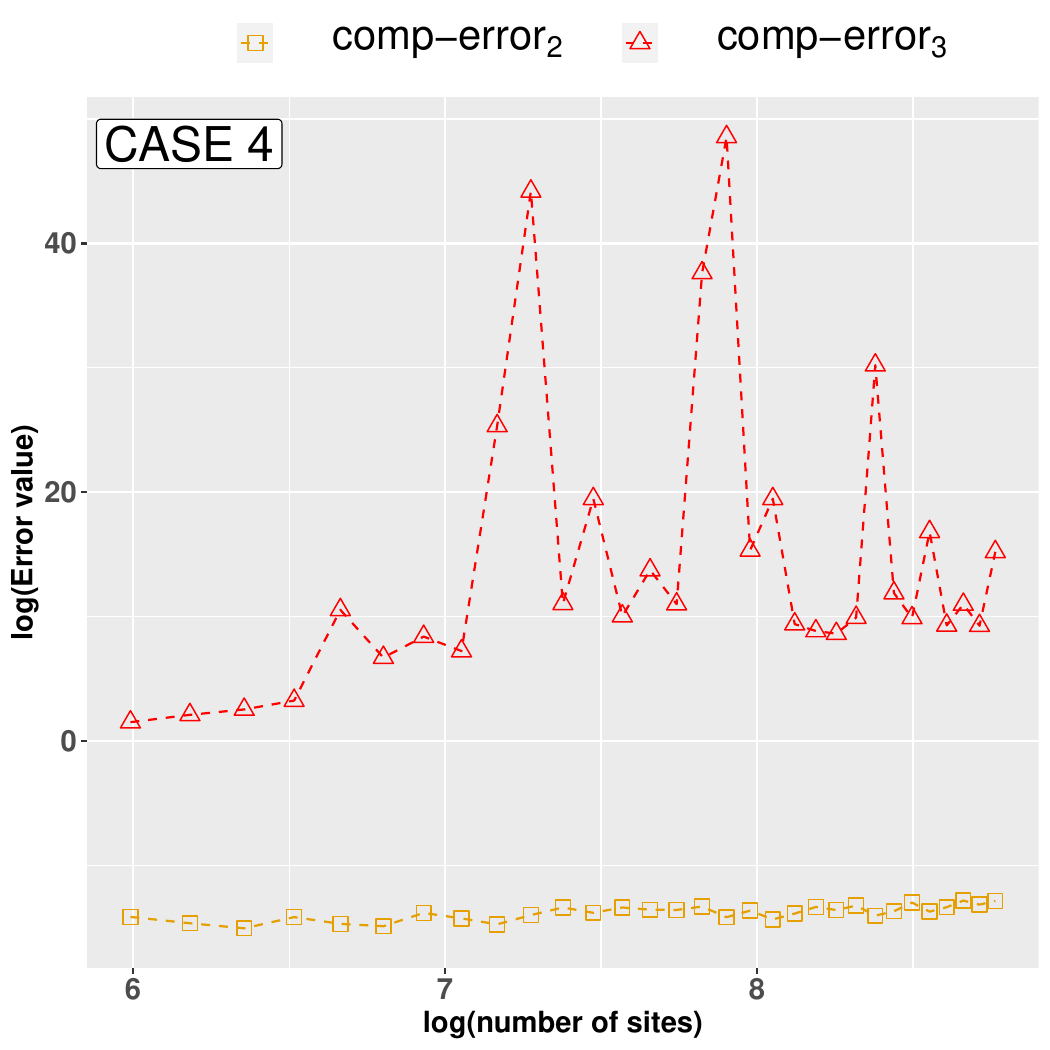}}
      \end{minipage}
\caption{\texttt{comp-error} for M2 and M3 compared with M1 using different parametrizations to obtain the $\mathcal{H}$-matrix}
\label{fig:plot7}
\end{figure}

We prescribed different values of the compression parameters $\varepsilon$ and $\eta$ and calculated \texttt{rmse} (Table \ref{table:table4}). We note that $\texttt{rmse}_j$ is almost constant if there is no compression present regarding the base case. However, the deviation
grows with lower degree of compression, i.e., higher tolerance $\varepsilon$ and/or admissibility parameter $\eta$. We also evaluated the computational performance (Figure \ref{fig:plot6}) and errors $\texttt{comp-error}_j$ (Figure \ref{fig:plot7}) for all four cases given in Table \ref{table:table4}.

\setlength{\tabcolsep}{1.5em} 
\begin{table}[!htb]
\centering
\caption{\texttt{rmse} for different compression parameters $\varepsilon$ and $\eta$ for interpolation purposes.}
\label{table:table4}\textbf{}
\begin{tabular}{ccc}
\hline
\textbf{Cases} &  parameters $\mathcal{H}$-matrix &  \texttt{rmse} \\
\hline
Case 1 & $\varepsilon = 0.01$,\hspace{2mm} $\eta = 5$  & 0.01  \\
\\
Case 2 & $\varepsilon = 0.01$, \hspace{2mm} $\eta = 10$ &  0.01    \\
\\
Case 3 & $\varepsilon = 0.1$, \hspace{2mm} $\eta = 5$  &   822  \\
\\
Case 4 & $\varepsilon = 0.1$, \hspace{2mm} $\eta = 10$ &   7526  \\
\hline
\end{tabular}
\end{table}
\addtolength{\tabcolsep}{1pt}

For cases 1 and 2, when the value of $\varepsilon$ is close to the value of case base ($\varepsilon = 0.0001$), the (\texttt{rmse}) is nearly the same. However, when this value is relaxed to expedite computation (cases 3 and 4), the \texttt{rmse} increases significantly, regardless of the value of $\eta$.

Since the smoothing parameter $\lambda$ controls the smoothness of the estimated function, various scenarios for different values of this parameter are presented for method M3 evaluating its influence in the interpolation. To this end, we employ $\texttt{rmse}$ as evaluation metric. As parameter $\varepsilon$ controls the accuracy of the compressed matrix, we changed only the $\lambda$ and $\eta$ values (Table \ref{table:table5}).
\vspace{0.5cm}

\begin{table}[!htb]
\centering
\caption{\texttt{rmse} for different values of $\lambda$ and $\eta$ in M3}
\label{table:table5}\textbf{}
\begin{tabular}{cccc}
\hline
 \text{Smoothing parameter} & $\varepsilon$        & $\eta$              & \texttt{rmse} \\
\hline
\\
$\lambda = 1$    & \multirow{4}{*}{0.0001}            & \multirow{4}{*}{2}        &  0.01    \\
$\lambda = 5$    &                   &                                            &  0.04    \\
$\lambda = 10$   &                   &                                            &  0.05    \\
$\lambda = 100$  &                   &                                            &  0.10     \\
\\
$\lambda = 1$    & \multirow{4}{*}{0.0001}            & \multirow{4}{*}{5}        &  0.01      \\
$\lambda = 5$    &                   &                                            &  0.04     \\
$\lambda = 10$   &                   &                                            &  0.05     \\
$\lambda = 100$  &                   &                                            &  0.10     \\
\\
$\lambda = 1$    & \multirow{4}{*}{0.0001}            & \multirow{4}{*}{10}       &  0.01      \\
$\lambda = 5$    &                   &                                            &  0.04     \\
$\lambda = 10$   &                   &                                            &  0.05     \\
$\lambda = 100$  &                   &                                            &  0.10      \\
\\
\hline
\end{tabular}
\end{table}
\vspace{0.5cm}

\subsection{Monte Carlo simulation}\label{sub4}

To evaluate the uncertainty in the approximation of the function $S(g)$, we perform a Monte Carlo simulation for spatial locations in a square grid of $20 \times 20$ distributed uniformly. Hence, the dense matrix is built in every iteration, and the Frank's function was perturbed adding an additional noise ($\varepsilon \sim N(0, 1)$) in every iteration as well. The grid size was selected to ascertain whether a limited number of spatial locations have implications in the estimated function of M3.

\begin{figure}[H]
  \centering
    \includegraphics[width=16cm, height=8cm]{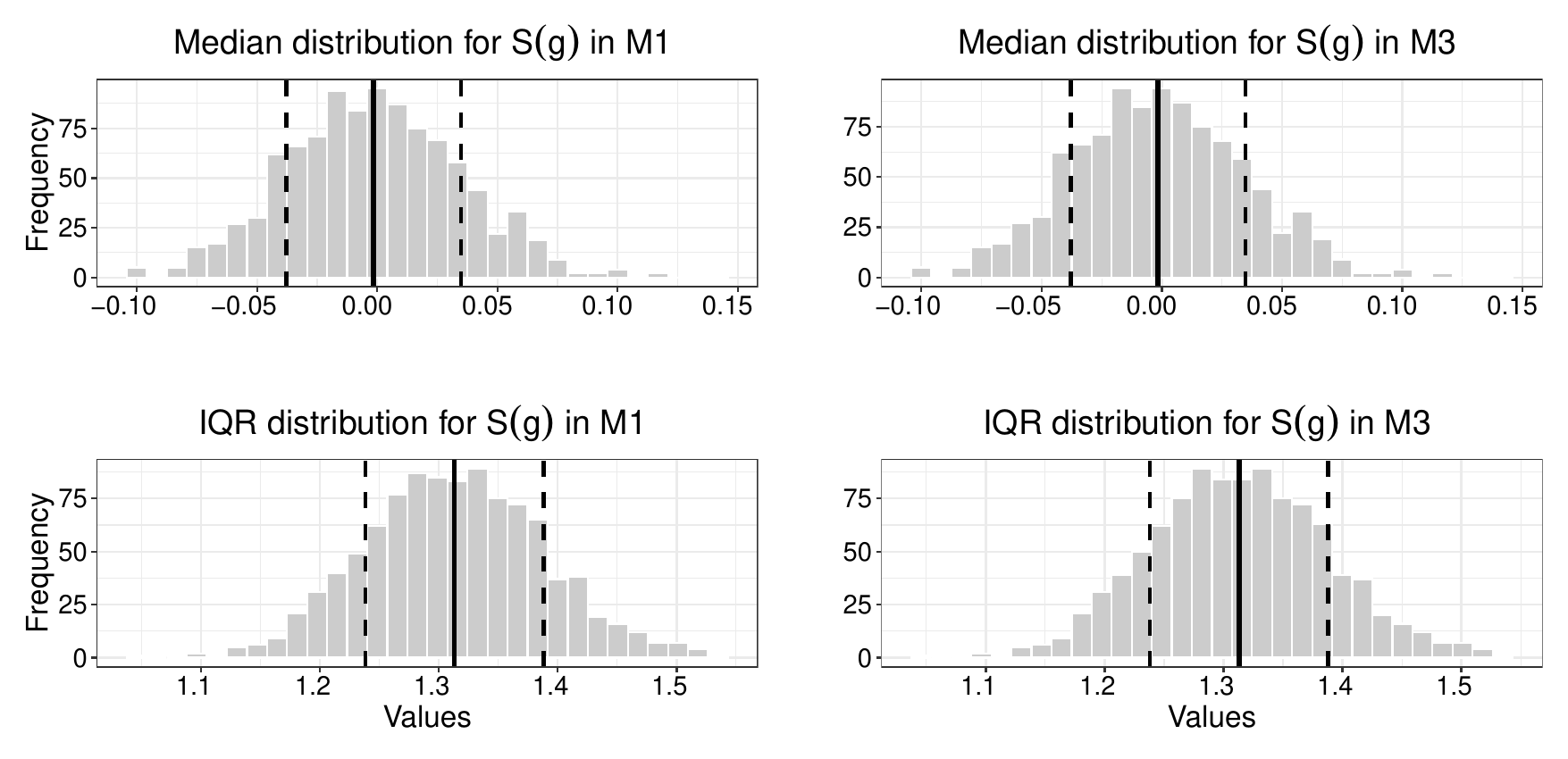}
\caption{Median (top) and IQR (bottom) distribution from the Monte Carlo simulations for M1 and M3 respectively. The solid lines represents the mean of the distribution and the dashed lines represents the standard deviations.} 
\label{fig:plot9}
\end{figure}

The Figure \ref{fig:plot9} shows the distribution of 1000 iterations using the Monte Carlo method for the sample median and sample interquartile range (IQR) for the function $S(g)$ from M1 and M3, along with their means (solid lines) and standard deviations (dashed lines). 

The distribution of the summary statistics (median and IQR) are similar between M1 and M3, this implies that M3 effectively approximates the function $S(g)$ when contrasted with M1 (exact method). Both M1 and M3 exhibit the same mean and standard deviation values for their respective distributions of the median, namely -0.001 and 0.036. The corresponding confidence interval is [-0.0038, 0.00072], lower and upper respectively. 

Finally, as these statistics (median and IQR) have a probability distribution, we measure their spreads by its variances (Table \ref{table:table6}),

\begin{table}[H]
\centering
\caption{Variances estimated for M1 and M3 from the samples of the Monte Carlo simulations.}
\label{table:table6}\textbf{}
\begin{tabular}{ccc}
\hline
 & Var(median) & Var(IQR) \\
\hline
M1 & 0.001  &  0.01\\
M3 & 0.001  &  0.01 \\
\hline
\end{tabular}
\end{table}

showing that the variances in the distributions of the median and IQR remain consistent between M1 and M3.

\newpage
\subsection{Additional comparisons}\label{sub5}

We also compare M3 with a thin plate spline regression (TPSR), which is available in the R package ``mgcv" (\cite{wood2011fast}). The methodology of TPSR is detailed in \cite{wood2003thin} and \cite{wood2017generalized}. For the TPSR, the \textit{knots} were chosen considering the grid itself. That is, we use 20 \textit{knots} for the grid of dimension $20 \times 20$  and so on to consider 80 \textit{knots} in the grid of $80 \times 80$. 

\begin{figure}[H]
  \centering
    \includegraphics[width=9cm, height=8.5cm]{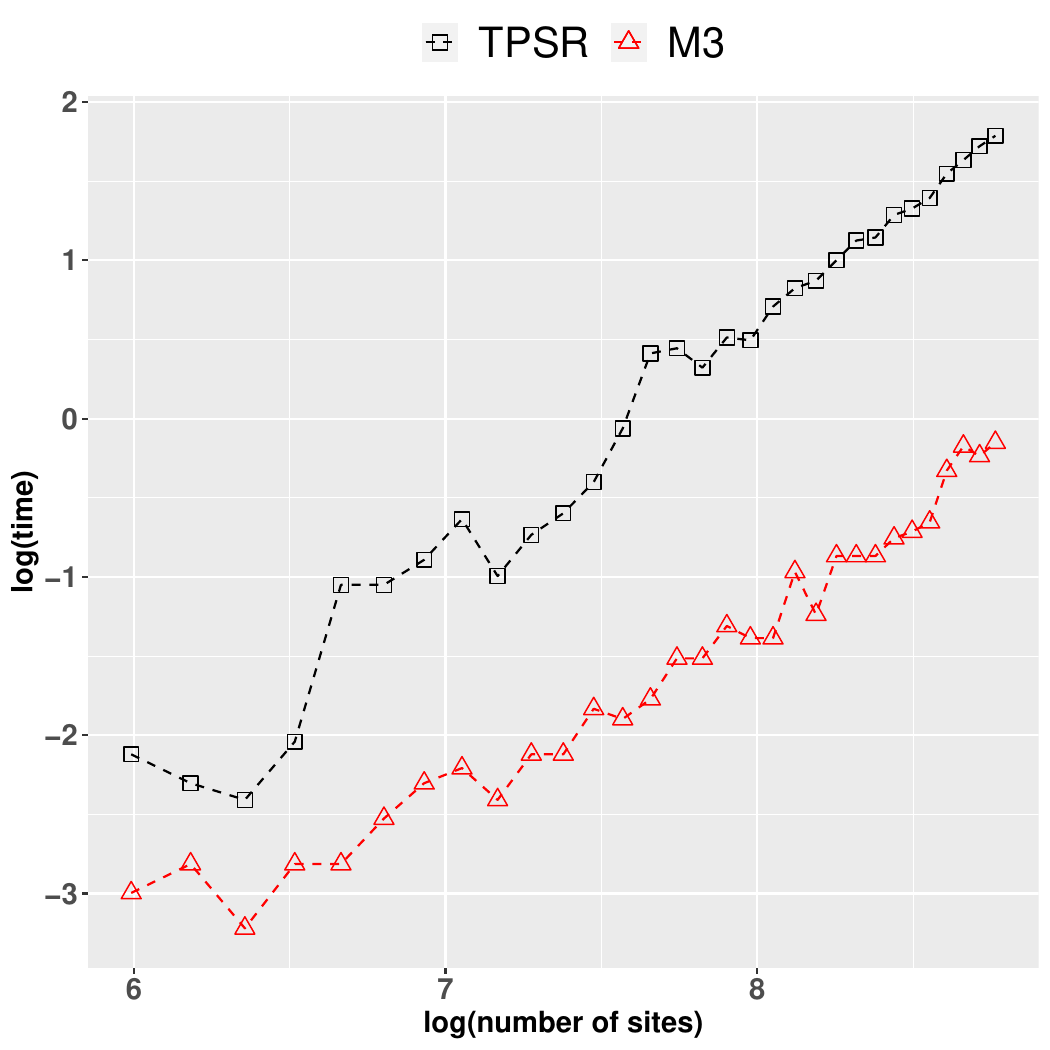}
\caption{Time of estimation for the TPRS and M3 computed in a sequences of grids from $20^{2}$ to $80^{2}$, represented on logarithmic scales.} 
\label{fig:plot10}
\end{figure}

The first comparison involves assessing the computational time required for the TPSR and M3 (Figure \ref{fig:plot10}).

\begin{figure}[H]
  \centering
    \includegraphics[width=12cm, height=6.5cm]{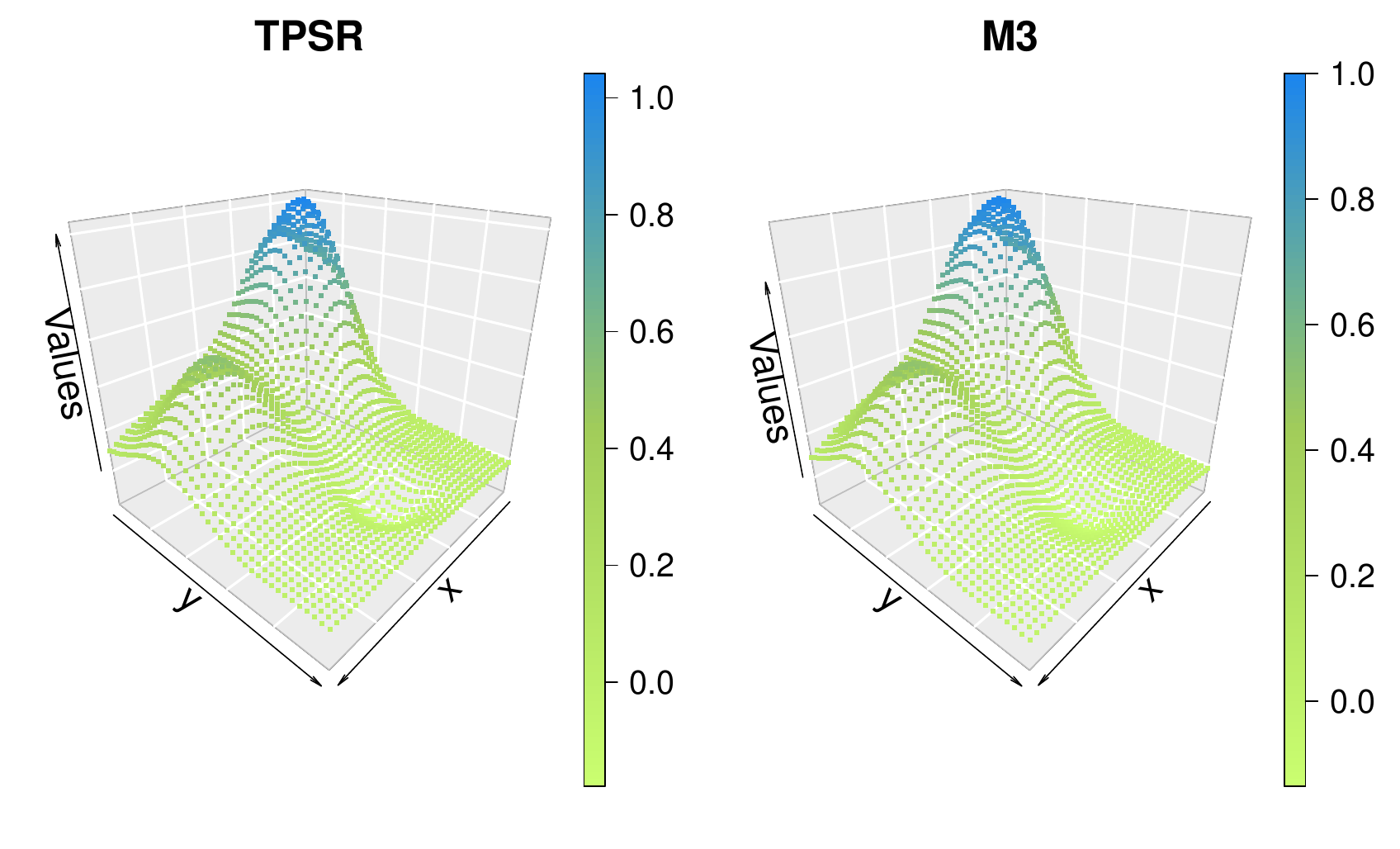}
\caption{Spatial interpolation in a grid of size $40^2$ from the TPSR and M3 respectively.} 
\label{fig:plot11}
\end{figure}

For this particular case, the plot show us that M3 outperforms TPSR in terms of computational time for the selected grids. Finally, we interpolate a function on a new grid with dimensions $40^{2}$ using method M3 and TPSR (Figure \ref{fig:plot11}).  

\begin{figure}[H]
  \centering
    \includegraphics[width=14cm, height=6.8cm]{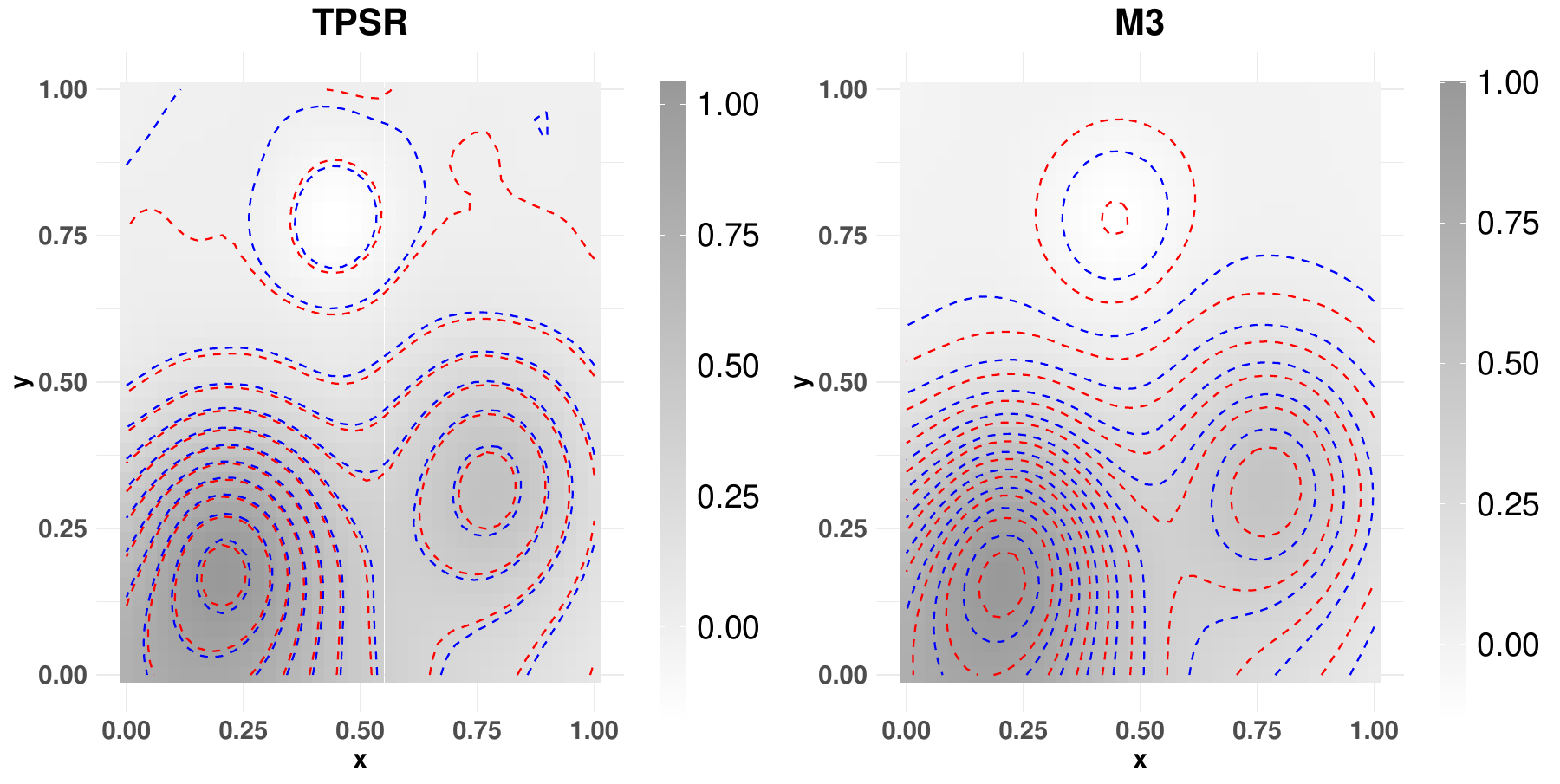}
\caption{Spatial interpolation in a grid of size $40^2$ from the TPSR and M3 respectively. In both cases, the dotted lines represents the lower confidence interval and the upper confidence interval.} 
\label{fig:plot12}
\end{figure}

The function interpolated by TPSR and M3 are quite similar, with the only distinction being that in the former, values tend to be higher based on the structure of the function on the new grid. Figure \ref{fig:plot12} shows the interpolation for both methods, alongside their corresponding confidence intervals. We can see that TPSR interpolation displays heightened uncertainty towards the limits of the new grid, as it attempts to interpolate values right up to its spatial domain boundaries. Conversely, M3 demonstrates a smoother behavior, refraining from interpolating values at the extreme ends of the spatial domain.
 
For the above, and considering the \texttt{rmse}, M3 demonstrates a similar statistical performance than TPSR (Table \ref{table:table7}).

\begin{table}[H]
\centering
\caption{\texttt{rmse} for the interpolation of the TPSR and M3 in a grid of size $40^{2}$.}
\label{table:table7}\textbf{}
\begin{tabular}{ccc}
\hline
 & TPSR & M3 \\
\hline
\texttt{rmse} & 0.004  &  0.010\\
\hline
\end{tabular}
\end{table}

\section{Discussion}
\label{section4}

We presented a fast method for computing a STPS for spatial data, based on an $\mathcal{H}$-matrix. Different simulations on distinct grid sizes, comparing the computational cost (in time) and the error of estimation for three different methods were presented. In this comparative, our proposal (M3) was most efficient in computational time and with an acceptable error compared with the exact method (M1). Additional scenarios with a small number of observations were considered for M3, and in all of them the approximations were similar to the results obtained initially, that is, a fast computation and a small error. Furthermore, Monte Carlo simulations were conducted to assess the uncertainty of the approximated function when facing perturbations in the response variable and spatial locations. Finally, a comparison between M3 and the TPSR revealed that, in the specific case of simulation, data interpolation using M3 exhibits nearly equivalent statistical performance to TPSR.

While recent experiments have explored similar approaches (\cite{litvinenko2019likelihood, chen2021linear}), it's important to note that these studies have primarily been conducted under the assumption of an underlying spatial stochastic process, which is itself determined by specific parameters defining its correlation structure. Furthermore, the estimation process was carried out utilizing the maximum likelihood method. Radial basis functions are not commonly employed directly in spatial or spatiotemporal modeling; however, they have been utilized to approximate covariance matrices within the probabilistic approach (\cite{lindgren2011explicit, porcu2013radial, nychka2015multiresolution}).



One disadvantage of using SPTS for spatial interpolation is that it does not explicitly assume a spatial correlation between the sites (locations), in contrast to Kriging, which utilizes variogram to describe the degree of spatial dependence of a spatial random field (\cite{cressie1989geostatistics}, \cite{oliver2015basic}). For the above, it becomes challenging to ascertain the distance at which the decay of spatial correlation initiates. However, it's worth noting that this aspect could also be considered a disadvantage for Kriging. This is because the effectiveness of spatial interpolation using Kriging relies on an underlying model (variogram) which, in turn, must be fitted using different correlation functions. On the other hand, STPS exhibit flexibility in accommodating spatial locations without necessitating the stringent assumption of stationarity, as the Kriging assumes (\cite{o1997spatial}). Furthermore, the inclusion of a ``smooth" component in its structure enables us to assume the absence of singularities among the data points. This is a similar assumption made in Kriging by introducing the nugget effect (\cite{hutchinson1994splines}). Regarding uncertainty in interpolation, Kriging provides a direct estimation of uncertainty through the variance, a quantity that STPS does not calculate. The variance in \ref{eq:smooth} can be estimated subsequent to fitting the spline model through  Generalized Cross Validation (GCV) method. In contrast, when applying Kriging, this variance is estimated prior to the computation of the surface (\cite{hutchinson1994splines}). Nevertheless, it is possible to derive this uncertainty using Monte Carlo simulations, as demonstrated in the study as well. In general, which method is more ``appropriate" depends on the application and the objectives of the analysis (\cite{handcock1994kriging}, \cite{ren2016comparing}). 

The current work also allows us to research in other mathematical methodologies to compress dense matrices, for example, using domain decomposition methods for scattered data interpolation (\cite{le2019domain}) or using a preconditioner to solve an augmented linear system but viewed from a probabilistic perspective (\cite{hennig2015probabilistic, oates2019modern, cockayne2019bayesian}). Also we think that  this is the first step to use an $\mathcal{H}$-matrix in the \texttt{R} software since it is also available in \texttt{Matlab} (\cite{ho2020flam}), \texttt{Python} (\cite{mikhalev2016iterative}) or \texttt{Julia} (\cite{geoga2020scalable}) as well. Today to use an \texttt{R} code with an $\mathcal{H}$-matrix is tricky, but all the machinery could be converted in a package and be available for the users in the future,  since \texttt{R} is an essential software for statistical analysis.


\section*{Declaration of Competing Interest}
The authors declare that he has no known competing financial interests or conflict of interest that could have appeared to influence the work reported in this paper.

\section*{CRediT authorship contribution statement}
{\bf Joaquin Cavieres:} Conceptualization, Formal analysis, Writing - original draft, Writing - review \& editing, Visualization.
{\bf Michael Karkulik:} Conceptualization, Formal analysis, Writing - original draft, Supervision, Writing - review \& editing.



\bibliographystyle{plainnat}
\bibliography{main}

\newpage

\appendix
\section{Appendix A}
\label{section5}

We aim to solve the augmented linear system expressed in Equation \eqref{eq:13}.
The matrix on the left-hand side of this system is not positive definite, and neither is the sub-matrix $\boldsymbol{E} + \lambda\boldsymbol{I}$.
However, $\boldsymbol{E} + \lambda\boldsymbol{I}$ is positive definite in the subspace
$\{ \boldsymbol{c}\in\mathbb{R}^n \mid \boldsymbol{P}^\top \boldsymbol{c} = \boldsymbol{0} \}$, since in this case

\begin{equation}\label{eq:14}
\boldsymbol{c}^{T}(\boldsymbol{E} + \lambda\boldsymbol{I})\boldsymbol{c} = \boldsymbol{c}^{T}\boldsymbol{E}\boldsymbol{c} + \lambda||\boldsymbol{c}||_2^{2} = J(g) + \lambda||\boldsymbol{c}||^{2}  \geq  \lambda||\boldsymbol{c}||^{2}.
\end{equation}

We separate the set of sites $S = \{ \boldsymbol{x}_1,\dots,\boldsymbol{x}_n\}$ into two sets
$S_{1} = \{\boldsymbol{x}_{i}, i = 1,..., n - 3\}$ and $S_{2} = \{\boldsymbol{x}_{n-2},\boldsymbol{x}_{n-1},\boldsymbol{x}_n \}$.
Without loss of generality, we can assume that the points in $S_{2}$ are not collinear.
We write accordingly $\boldsymbol{c} = (\boldsymbol{c}_{1}, \boldsymbol{c}_{2})$ and
$\boldsymbol{y} = (\boldsymbol{y}_{1}, \boldsymbol{y}_{2})$ and the linear system in Equation \eqref{eq:13} becomes

\begin{align}\label{eq:15}
  \begin{pmatrix}
    \boldsymbol{E}_{11} + \lambda\boldsymbol{I} & \boldsymbol{E}_{12} & \boldsymbol{P}_{1}\\
    \boldsymbol{E}_{21} & \boldsymbol{E}_{22} + \lambda\boldsymbol{I} & \boldsymbol{P}_{2} \\
    \boldsymbol{P}^\top_{1} & \boldsymbol{P}^\top_{2} & \boldsymbol{0}
  \end{pmatrix}
  \begin{pmatrix}
    \boldsymbol{c}_{1} \\ \boldsymbol{c}_{2} \\ \boldsymbol{d}
  \end{pmatrix}
  =
  \begin{pmatrix}
    \boldsymbol{y}_{1} \\ \boldsymbol{y}_{2} \\ \boldsymbol{0}
  \end{pmatrix},
\end{align}

where $\boldsymbol{E}_{11} + \lambda\boldsymbol{I} \in \mathbb{R}^{(n-3)\times (n-3)}$ and
$\boldsymbol{E}_{22} + \lambda\boldsymbol{I} \in \mathbb{R}^{3 \times 3}$.
Note that $\boldsymbol{P}_{2}$ is invertible because the points in $S_{2}$ are not collinear, and so

\begin{equation*}
  \begin{pmatrix}
    \boldsymbol{E}_{22} + \lambda\boldsymbol{I} & \boldsymbol{P}_{2}\\
    \boldsymbol{P}_{2}^\top & \boldsymbol{0}
  \end{pmatrix}  
\end{equation*}
is invertible. Given the above, we can use the Schur complement to write

\begin{align}\label{eq:16}
\begin{split}
(\boldsymbol{E}_{11}+\lambda\boldsymbol{I})\boldsymbol{c}_{1} -
  \begin{pmatrix}
    \boldsymbol{E}_{12} & \boldsymbol{P}_{1}
  \end{pmatrix}
  &
   \begin{pmatrix}
    \boldsymbol{E}_{22}+\lambda\boldsymbol{I} & \boldsymbol{P}_{2}\\
    \boldsymbol{P}_{2}^\top & \boldsymbol{0}
  \end{pmatrix}^{-1}
   \begin{pmatrix}
    \boldsymbol{E}_{21}\\
    \boldsymbol{P}_{1}^\top
  \end{pmatrix} \boldsymbol{c}_{1}\\
  &= \boldsymbol{y}_{1} -   
  \begin{pmatrix}
    \boldsymbol{E}_{12} & \boldsymbol{P}_{1}
  \end{pmatrix} 
    \begin{pmatrix}
    \boldsymbol{E}_{22} + \lambda\boldsymbol{I} & \boldsymbol{P}_{2}\\
    \boldsymbol{P}_{2}^\top & \boldsymbol{0}
  \end{pmatrix}^{-1}
    \begin{pmatrix}
    \boldsymbol{y}_{2} \\ \boldsymbol{0}
  \end{pmatrix}
\end{split}
\end{align}

The matrix of this linear system

\begin{equation*}
\boldsymbol{M} =  
\boldsymbol{E}_{11} + \lambda\boldsymbol{I} -
\begin{pmatrix}
    \boldsymbol{E}_{12} & \boldsymbol{P}_{1}
  \end{pmatrix} 
   \begin{pmatrix}
    \boldsymbol{E}_{22} + \lambda\boldsymbol{I} & \boldsymbol{P}_{2}\\
    \boldsymbol{P}_{2}^\top & \boldsymbol{0}
  \end{pmatrix}^{-1}
   \begin{pmatrix}
    \boldsymbol{E}_{21}\\
    \boldsymbol{P}_{1}^\top
    \end{pmatrix}
\end{equation*}

is obviously symmetric. Furthermore, it is positive definite, since for $\boldsymbol{0}\neq\boldsymbol{c}_{1} \in \mathbb{R}^{n-3}$ we have

\begin{align*}
\boldsymbol{M}\boldsymbol{c}_{1} &=  
(\boldsymbol{E}_{11}+\lambda\boldsymbol{I})\boldsymbol{c}_{1} +
\begin{pmatrix}
    \boldsymbol{E}_{12} & \boldsymbol{P}_{1}
  \end{pmatrix} 
   \begin{pmatrix}
    \boldsymbol{E}_{22} +\lambda\boldsymbol{I}& \boldsymbol{P}_{2}\\
    \boldsymbol{P}_{2}^\top & \boldsymbol{0}
  \end{pmatrix}^{-1}
   \begin{pmatrix}
    -\boldsymbol{E}_{21}\boldsymbol{c}_{1} \\
    -\boldsymbol{P}_{1}^\top \boldsymbol{c}_{1}   
    \end{pmatrix} \\
&=
(\boldsymbol{E}_{11}+\lambda\boldsymbol{I})\boldsymbol{c}_{1} +  
\begin{pmatrix}
    \boldsymbol{E}_{12} & \boldsymbol{P}_{1}
  \end{pmatrix} 
\begin{pmatrix}
    -\boldsymbol{P}^{-\top}_{2}\boldsymbol{P}_{1}^\top\boldsymbol{c}_{1} \\
     -\boldsymbol{P}^{-1}_{2}\boldsymbol{E}_{21}\boldsymbol{c}_{1} - \boldsymbol{P}^{-1}_{2}(\boldsymbol{E}_{22}+\lambda\boldsymbol{I})
      (-\boldsymbol{P}^{-\top}_{2}\boldsymbol{P}_{1}^\top)\boldsymbol{c}_{1}
    \end{pmatrix} \\
&=
(\boldsymbol{E}_{11}+\lambda\boldsymbol{I})\boldsymbol{c}_{1} + 
\boldsymbol{E}_{12}(-\boldsymbol{P}^{-\top}_{2}\boldsymbol{P}_{1}^\top)\boldsymbol{c}_{1} + 
(-\boldsymbol{P}_{1}\boldsymbol{P}^{-1}_{2})\boldsymbol{E}_{21}\boldsymbol{c}_{1} + (-\boldsymbol{P}_{1}\boldsymbol{P}^{-1}_{2})(\boldsymbol{E}_{22}+
\lambda\boldsymbol{I})\boldsymbol{P}^{-\top}_{2}\boldsymbol{P}_{1}^\top\boldsymbol{c}_{1}
\end{align*}

We conclude that

\begin{equation*}
\boldsymbol{c}^\top_{1}\boldsymbol{M}\boldsymbol{c}_{1} = 
\begin{pmatrix}
    \boldsymbol{c}_{1} \\
    -\boldsymbol{P}^{-\top}_{2}\boldsymbol{P}_{1}^\top\boldsymbol{c}_{1}
  \end{pmatrix}^\top (\boldsymbol{E} + \lambda\boldsymbol{I})
  \begin{pmatrix}
    \boldsymbol{c}_{1} \\
   -\boldsymbol{P}^{-\top}_{2}\boldsymbol{P}_{1}^\top\boldsymbol{c}_{1}
  \end{pmatrix},
\end{equation*}

and hence $\boldsymbol{M}$ is positive definite since Equation \eqref{eq:14},

\begin{equation*}
\boldsymbol{P}^\top\begin{pmatrix}
    \boldsymbol{c}_{1} \\
    -\boldsymbol{P}^{-\top}_{2}\boldsymbol{P}_{1}^\top\boldsymbol{c}_{1}
  \end{pmatrix} = \boldsymbol{0}.
\end{equation*}

\end{document}